\documentstyle[pre,aps,epsfig,preprint]{revtex}
\newcommand{\qvec}{{\bf q}}
\newcommand{\kvec}{{\bf k}}
\newcommand{\rvec}{{\bf r}}

\newcommand{\beq}{\begin{equation}}
\newcommand{\eeq}{\end{equation}}
\newcommand{\beqa}{\begin{eqnarray}}
\newcommand{\eeqa}{\end{eqnarray}}

\tightenlines
\begin{document}    
\draft
\title{Excluded volume effects on the structure of a linear polymer 
under shear 
flow}
\author{Carlo Pierleoni}
\address{INFM and Dipartimento di Fisica, Universit\`a degli Studi,
Via Vetoio, I-67100 
L'Aquila (Italy), \\email: carlo.pierleoni@aquila.infn.it}
\author{Jean-Paul Ryckaert}
\address{D\'epartement de Physique, Universit\'e Libre de Bruxelles, 
CP223
Brussels 
(Belgium)\\ email: jryckaer@ulb.ac.be} 
\date{\today}
\maketitle
\begin{abstract}
The effect of excluded volume interactions on the structure of 
a polymer in shear flow is investigated by Brownian Dynamics 
simulations for chains with size $30\leq N\leq 300$. The main 
results concern the structure factor $S({\bf q})$ of chains of N=300 Kuhn segments, 
observed  at a reduced shear rate 
$\beta=\dot{\gamma}\tau=3.2$, where $\dot{\gamma}$ is the bare shear rate and 
$\tau$ is the longest relaxation time 
of the chain. At low q, where anisotropic global deformation is 
probed, the chain form factor 
is shown to match the form factor 
of the continuous Rouse model under shear at the same reduced shear rate,
computed here for the first time in a wide range of wave vectors. At high q, the 
chain structure factor evolves towards the isotropic equilibrium  
power law $q^{-1/\nu}$ typical of self-avoiding walk statistics.
The matching between excluded volume and ideal chains at small q, 
and the excluded volume power law behavior at large q are 
observed for ${\bf q}$ orthogonal to the main 
elongation axis but not yet for ${\bf q}$ along the elongation 
direction itself, as a result of interferences with finite extensibility  
effects. Our simulations support the existence of anisotropic shear 
blobs for polymers in good solvent under shear flow for $\beta>1$ 
provided chains are sufficiently long.
\end{abstract}


\newpage
\narrowtext

\section{Introduction}

Excluded volume (EV) and hydrodynamics interactions (HI) have been recognized as 
fundamental physical phenomena to explain the peculiarities of single chain 
behavior in dilute polymer solutions. 
For a solution at rest, their effects on the single chain structure and dynamics 
have been carefully investigated over the years both theoretically and 
experimentally\cite{Y71,Ge79,DE86,dCJ90}.
When a dilute solution is subjected to a flow field, the situation is much more 
complex. In addition to finite extensibility (FE) effects to be considered at 
high strain, the role of EV and HI effects is  still largely debated. This 
situation certainly arises from the difficulties to solve the kinetic equations 
for the dynamics of the simple coarse-grained models
when EV and HI are considered. 
Some progresses have been made concerning the HI effects in absence of EV 
interactions by the use of a self consistent preaveraging approximations 
\cite{O96}, a non equilibrium version of the more famous equilibrium Zimm 
theory\cite{DE86}.
Renormalization Group Theory techniques have also been applied to 
solve suitable dynamic equations for the chain 
\cite{PSO86,W89,O90,W90,ZO91,BH90,RK89}.
The main focus of those investigations was the rheological response of the 
system although some results on the chain structure where also 
provided\cite{ZO91,W90}. An attempt to introduce EV effects in the Rouse model 
has also been performed by perturbation theory expansion\cite{Carl95-96}.

On the experimental side the situation is similar in that  measurements of 
single chain properties in flow are considerably more difficult than at rest. 
However, several scattering experiments have been performed, mainly Light 
Scattering (LS), with the aim to investigate the global properties of the chain 
such as the orientation and the deformation of the coil by the flow 
\cite{CMS69,LO88,Li91,LS93,ZS94-95,LSM97,LM99}.
The effect of EV interaction over global chain properties has been discussed
in a series of recent LS studies \cite{ZS94-95,LSM97,LM99} where some systematic 
trends were reported. For dilute solutions in shear flow at a fixed 
reduced shear rate, a systematic weakening of the alignment of the main axis of 
the chain 
in the flow direction has been observed for increasing solvent quality. 
At a given reduced shear rate, the measured 
expansion ratio indicates that good solvent chains are less deformed than theta 
chains  \cite{LM99}.
Inspection of the flow effects on the internal structure of the coil can only
be obtained by Small Angle Neutron Scattering (SANS) and indeed some data have 
been already published\cite{LO88}. However experimental limitations are here 
even more severe than for LS, and a quantitative study for various solvent 
qualities is still missing.

New experimental insight in the single chain behavior under flows and stretching 
forces has been provided recently by fluorescence techniques on biological 
molecules like DNA 
\cite{BSF92,BMSS94,CLHLVCC96,ROH97,PSC94,PSLC95,PQDC94,PSC97,SBC99,LHBW99}. 
Two papers on the single chain dynamics under shear flow have been published 
recently by S. Chu and coworkers \cite{SBC99} and by D. Wirtz and coworkers 
\cite{LHBW99} in both cases employing DNA chains. In the former paper,the  mean 
and time fluctuations of the fractional extension of DNA under shear flow in the 
flow direction was reported  for reduced shear rates larger than one and 
culminating at $\beta=76$. In the second paper, the authors have shown that the 
dynamics of a DNA chain under weak shear flow ($\beta\sim 0.1$) presents an 
unexpected  wide range of relaxation 
times related to the particular initial configuration prior to the flow inset.

An interesting alternative to the experimental tools in the context of polymer 
physics are the numerical techniques such as Monte Carlo (MC), Molecular 
Dynamics (MD) and Brownian Dynamics (BD) simulations. They can be used to solve 
simple models for the system of interest and therefore test or suggest 
theoretical developments or stimulate new experiments. On the other hand 
simulations can help interpreting experimental data in a more complete and 
coherent fashion. 

Many numerical studies of the behavior of a single chain under shear flow have 
already appeared by various techniques such as 
BD \cite{Liu89,Carl,Liulyn,ADMH98,RKKZ99,LL00}, 
MD \cite{PR93,PR95,RP99,AKH99} and even non-equilibrium MC \cite{DR,Lai}. To 
mention a recent simulation work \cite{LL00}, BD studies for a chain without HI 
and EV found a reasonable agreement with the infinite chain Rouse model 
predictions but came to the disturbing conclusion that, while the relative 
elongation of DNA chain in the flow direction is correctly explained by the 
chain model, the same model fails to explain the global expansion ratio of a 
polystyrene chain which is experimentally much lower than predicted. 
 
To our knowledge, the role of the excluded volume effect on the structure of the 
chain under SF has not been systematically studied by simulations. In the 
present paper we try to elucidate this effect by comparing the analytical 
results of the Rouse model (in the continuum limit) with BD data for a suitable 
microscopic model of flexible chain with excluded volume interactions and 
unavoidable FE features.

The structure of a single chain in shear flow is often discussed by exploiting 
an analogy with the well known structure of a single chain subjected to a pair 
of equal and opposite stretching forces applied to its ends. In this case the 
"macroscopic" behavior of long chains, as measured by the elasticity law, is 
characterized by a linear hookean regime for weak forces followed by a non 
linear power law regime for strong forces and ultimately by specific finite 
extensibility effects. The crossover between the first two regimes 
is linked to the appearance of the so called ``tensile'' blobs in the chain 
structure\cite{P76,Ge79}. 

A very clear signature of the blobs can be 
detected in the chain structure factor as a crossover from ideal behavior at 
small q's to excluded volume statistics at large q's.
This has been shown long time ago by Neutron Scattering
experiments for ``thermal'' blobs in the semi-diluted solutions at rest 
\cite{FBC78}, and recently by MC simulation for ``tensile'' blobs in the 
stretched chain problem\cite{PAR97}. In the latter case, the crossover is 
detected for q orthogonal to the external force and can be ascribed to the fact 
that scattering at small q's in these directions comes from pairs of monomers 
which are far apart in the elongation direction and therefore do not interact 
directly through the EV potential.
 
For the chain in shear flow Onuki has formulated a blob model\cite{On85}
as a simple extension of the Pincus-deGennes theory for the stretched chain 
case\cite{P76,Ge79}. The blob size is related to the bare shear rate and the 
chain at high shear rates is imagined as a string of shear blobs aligned most of 
the 
time in the flow direction. Such a structure which corresponds to an aligned 
fluctuating rod, should yield a scattering signal for $q$ in the flow direction 
(flow and elongation directions provide essentially the same scattering signal 
at high enough shear rate\cite{PR95}) similar to the stretched chain signal for 
$q$ aligned in the stretching direction \cite{PAR97}. For the stretched chain 
such behavior is described quantitatively by the Rouse model with a suitable 
choice of the chain elongation and the longitudinal and transverse fluctuations 
\cite{PAR97,BDODCFJ75}. In the present paper we test these ideas on the form 
factor of a single EV chain in flow for shear rates which are sufficiently high 
to reach the plausible shear blob regime but also sufficiently small to remain 
in the scaling regime, i.e. avoiding finite chain effects. Our strategy is to 
analyze to which extend the low q behavior resembles the 
single chain structure factor of ideal chains in the presence of shear flow, 
which can be derived exactly, and whether the high q regime reproduces the 
structure of unperturbed  EV chains. In order to observe such phenomena taking 
place 
on different length scales, we need sufficiently long chains and therefore we 
have performed BD simulations of EV chains in absence and presence of shear 
flow.

The paper is organized as follows. In section \ref{sec:phenom} we set up the 
essential phenomenology concerning a chain in shear flow and define the relevant 
structural characteristics of the chain. Section \ref{sec:rouse} describes the 
procedure to solve the Rouse model in flow and provides the results for the 
specific SF case. This is not a totally new analytical development as we are 
aware of at least two previous works on it \cite{PB88,Carl}. The novelty here is 
however the calculation of the structure factor in the whole range of $q$ 
vectors and the use of such behavior to pinpoint the signature of shear blobs in
the EV chain structure factor.
In the next section \ref{sec:bd} we describe the model for EV chains and the 
BD technique used. 
In section \ref{sec:results} we discuss the results of BD simulations. 
Finally section \ref{sec:conclusions} is devoted to our conclusion and 
perspectives.

\section{Phenomenological framework}\label{sec:phenom}

When a dilute polymer solution is subjected to a simple shear flow,
characterized by the velocity field
${\bf u}=\dot\gamma y {\bf \hat x}$ where$\dot\gamma$ is the shear rate, 
the chains in the solution are oriented and 
deformed
according to the flow. Those phenomena can be measured through the anisotropy
arising in any tensorial quantity associated to the chain, such as for instance 
the gyration tensor ${\bf G}$ and the order parameter tensor ${\bf O}$ defined 
as
\beqa
{\bf G}&=&\frac{1}{2N^2}\sum_{i,j=1}^N <({\bf R}_i-{\bf R}_j)({\bf R}_i-{\bf 
R}_j)>\\
{\bf O}&=&\sum_{i=1}^{N-1}\left<\frac{{\bf u}_i{\bf u}_i}{|{\bf u}|^2} - 
\frac{{\bf 1}}{3}\right>
\eeqa
where ${\bf R}_i$ are the coordinates of the i-th monomer in the chain, 
${\bf u}_i={\bf R}_{i+1}-{\bf R}_i$
and ${\bf 1}$ is the unit tensor. 
At equilibrium the system is isotropic and any tensorial quantity, say ${\bf A}$ 
reduces to 
a scalar quantity. Under shear flow, symmetry imposes 
$A_{xz}=A_{zx}=A_{yz}=A_{zy}=0$. 
Therefore there are at most four independent components of any tensor to be 
monitored,
namely the three diagonal elements and the off-diagonal $A_{xy}=A_{yx}$. 
The orientational angle $\chi_A$, defined through the relation
\beq
cot(2\chi_A)=\frac{A_{xx}-A_{yy}}{2A_{xy}}
\label{eq:chi}
\eeq
measure the rotation around ${\bf \hat z}$ of the principal axes (I,II,III) of 
the tensor 
${\bf A}$
with respect to the flow axes (x,y,z). In shear flow, $A_{xy}$ starts linearly 
with $\dot\gamma$ while the first contribution to $A_{xx}-A_{yy}$ is of order 
$\dot\gamma^2$.
Therefore the linear (Newtonian) regime is characterized by $\chi_A=\pi/4$. 
Outside the linear
regime $\chi_A$ decreases to zero for increasing shear rate.

The deformation ratios, defined by
\beq
\delta A_{\alpha\alpha}=\frac{A_{\alpha\alpha}(\beta)}{A_{\alpha\alpha}(0)}-1 
\eeq
measures the amount of deformation of the chain in flow, either in the flow 
reference frame 
($\alpha=x,y,z$), or in the molecular reference frame ($\alpha=I,II,III$). 
When defined in the flow reference frame, the first 
contribution to those quantities is of order $\dot\gamma^2$, 
so that the Newtonian regime is characterized
by the absence of deformation.
Outside the linear regime, the chain is elongated in the flow direction ($\delta 
A_{xx}>0$) and 
compressed in the in-plane gradient direction ($\delta A_{yy}<0$), while the 
neutral (out of plane)
direction is only slightly decreased ($\delta A_{zz}\sim 0$).

\section{Structure factor of the Rouse chain under shear flow}\label{sec:rouse}

To compute the structure factor of an ideal gaussian chain (Rouse chain) we 
adopt 
the continuous chain model introduced by Edwards\cite{DE86} in which the only 
energy term is a quadratic potential acting between nearest neighbors along the 
chain.
For such a model the equilibrium probability distribution in configurational 
space is
\beq
\Psi[{\bf R}]\propto exp\left(-\frac{\chi}{2}
\int_0^N dn \left(\frac{\partial {\bf R}_n}{\partial n}\right)^2 \right)
\eeq
where $\chi=3k_BT/d^2$, ${\bf R}_n$ represents the positions of the chain 
monomers, 
$0\leq n\leq N$ is the continuous monomer index and $d$ is the single bond 
variance 
of the model.
The steady state distribution in presence of a generic homogeneous solvent flow 
field can be obtained in terms of 
the normal modes of the model $({\bf X}_p, (p=0,\infty))$\cite{DE86,BCAH87}.
For homogeneous flows to which we limit our discussion, 
the normal modes are defined by the same transformation as at equilibrium. 
In those new coordinates the distribution is factorized 
over the normal modes 
\beq
\hat\Psi(\{{\bf X}\},t)=\prod_{p=0}^\infty \hat\psi_p({\bf X}_p,t)
\label{eq:PsiX}
\eeq
In SF, the steady state distribution of the {\it p-th} mode is
\beq
\hat\psi_p({\bf X}_p)=\left(\frac{\chi_1}{2\pi k_BT}\right)^{3/2} 
\frac{p^3}{\sqrt{1+\dot\gamma^2\left(\frac{\tau_1}{2p^2}\right)^2}}
e^{ -\frac{\chi_1}{2k_BT}p^2{\bf X}_p\cdot\beta_p^{-1}\cdot{\bf X}_p }
\label{eq:psipf}
\eeq
where the matrix $\beta_p$ is defined as
\beq
\beta_p={\bf 1}+\frac{\tau_1}{2p^2}(\kvec+\kvec^T)
+2\left(\frac{\tau_1}{2p^2}\right)^2\kvec\cdot\kvec^T
\label{eq:betap}
\eeq
and ${\bf k=\nabla u}$ is the velocity gradient tensor.
The derivation of eqs. (\ref{eq:psipf}) and (\ref{eq:betap}) is given in 
appendix \ref{app:ssd}.

The chain structure factor is defined as
\beq
S(\qvec,N,\dot\gamma)=\frac{1}{N}\int_0^N\int_0^N dn dm 
<e^{i\qvec\cdot{\bf R}_{nm}}>_{\dot\gamma}
\eeq
where ${\bf R}_{nm}={\bf R}_{n}-{\bf R}_m$ and $<...>_{\dot\gamma}$ indicates a 
statistical 
averages over the non equilibrium distribution eq.(\ref{eq:PsiX}). 
In terms of normal modes we have 
\beq
{\bf R}_{nm}=2\sum_{p=1}^\infty \left(cos\frac{p\pi n}{N}-cos\frac{p\pi 
m}{N}\right){\bf X}_p
\eeq
and using the model solution eq.(\ref{eq:psipf}) we obtain, after a gaussian 
integration,
\beq
<e^{i\qvec\cdot{\bf R}_{nm}}>_{\dot\gamma}=
exp\left\{
-\frac{2k_BT}{\chi_1} \qvec\cdot\left[\sum_{p=1}^\infty\frac{\beta_p}{p^2}
\left(cos\frac{p\pi n}{N}-cos\frac{p\pi m}{N}\right)^2\right]\cdot\qvec
\right\}
\eeq
Substituting the expression of $\beta_p$ from eq.(\ref{eq:betap}) we get
\beq
<e^{i\qvec\cdot{\bf R}_{nm}}>_{\dot\gamma}=
exp\left\{-\frac{2k_BT}{\chi_1} 
\left[
S_0(n,m) q^2 + 
\frac{\tau_1}{2} S_1(n,m) \qvec\cdot(\kvec+\kvec^T)\cdot\qvec +
\frac{\tau_1^2}{2} S_2(n,m) \qvec\cdot(\kvec\cdot\kvec^T)\cdot\qvec
\right]
\right\}
\label{eq:eiqrss}
\eeq
with 
\beq
S_k(n,m)=\sum_{p=1}^\infty\frac{1}{p^{2+2k}}
\left[cos\frac{p\pi n}{N}-cos\frac{p\pi m}{N}\right]^2 
\hskip 2cm k=0,1,2
\label{eq:sumsq}
\eeq
In eq.(\ref{eq:eiqrss}) the first term in the exponent on the r.h.s. is the 
equilibrium contribution
leading to the well known Debye function. The other two terms arise from the 
presence of the flow.
The three series above can be summed analytically as explained in appendix 
\ref{app:sums} and the 
result is
\beqa
S_0(a,b)&=&\frac{\pi^2}{2}|a-b| \\ 
S_1(a,b)&=&\frac{\pi^4}{4}(a-b)^2\left[(a+b)\left(1-\frac{a+b}{2}\right)-
\frac{|a-b|}{3}\right] \\ 
S_2(a,b)&=&\frac{\pi^6}{240}(a-b)^2\left\{|a-b|^3+5(a+b)\left[(a+b)(2+a^2+b^2)-
3a^2-2ab-3b^2\right]
\right\}
\eeqa
where we have defined reduced indices $a=n/N, b=m/N$ and $0\leq (a,b)\leq1$.

Let us now introduce the reduced shear rate $\beta$ as the product of the bare 
shear rate
and the global relaxation time of the chain at equilibrium. To compare the 
theory with 
the experiments we define $\beta$ as
\beq
\beta=\frac{M \eta_s [\eta] \dot\gamma}{N_A k_BT}
\eeq
where $M$ is the molecular weight of the polymer, $\eta_s$ the solvent shear 
viscosity,
$[\eta]$ the intrinsic viscosity of the solution and $N_A$ the Avogadro number.
In the Rouse model the intrinsic viscosity is 
$[\eta]=N_A/M\eta_s~(N^2d^2\xi/36)$ so that 
\beq
\beta=\frac{\dot\gamma}{k_BT}\frac{N^2d^2\xi}{36}
\eeq
which gives $\tau_1\dot\gamma=12\beta/\pi$\cite{DE86}. The chain structure 
factor in 
terms of the reduced shear rate is finally obtained as
\beq
\frac{S(\qvec,N,\dot\gamma)}{N}=\int_0^1da~\int_0^1 db~
e^{-\frac{Nd^2}{6}\qvec\qvec^T{\bf :f}(a,b,\beta)}
\label{eq:sqsf}
\eeq
where ${\bf f}(a,b,\beta)$ is the following tensor
\beq
{\bf f}(a,b,\beta)=|a-b|{\bf 1} +
\frac{12\beta}{\pi^4}\left[S_1(a,b)(\Lambda+\Lambda^T)+
\frac{12\beta}{\pi^2}S_2(a,b)(\Lambda\cdot\Lambda^T)\right]
\label{eq:ftensor}
\eeq
and $\Lambda=\dot\gamma^{-1}{\bf k}$.
Expressions (\ref{eq:sqsf}) with (\ref{eq:ftensor}) are our main interest here. 
The double integral over monomers 
cannot be performed analytically and we resort to numerical method.
Before presenting specific results we want to make some further considerations.
Firstly we note that the tensor ${\bf f}$ entering in the structure factor 
depends on 
$N$ and $\dot\gamma$ through $\beta$ only. This implies the following 
universality for the 
structure factor
\beq
\frac{S(\qvec,N,\dot\gamma)}{N}=F(R_g\qvec,\beta)
\eeq
where $F$ is a universal scalar function depending on the geometry of the 
applied flow 
and $R_g=\sqrt{Nd^2/6}$ is the equilibrium radius of gyration for the gaussian 
chain.
Secondly we want to obtain the well known expression for the gyration tensor of 
the ideal chain
under shear flow\cite{PB88,Carl} from which we can derive the orientation and 
deformation of the 
chain to be used below.
The small $q$ expansion to order $q^2$ of the structure factor is directly 
related to the 
gyration tensor by
\beq
\frac{S(\qvec,N,\dot\gamma)}{N} = 1 - \qvec\qvec^T {\bf :G}+O(q^4)
\eeq
When applied to eq.(\ref{eq:sqsf}) it provides
\beq
{\bf G}(N,\beta)=\frac{Nd^2}{6} \int_0^1da~\int_0^1 db~{\bf f}(a,b,\beta)=
\frac{Nd^2}{6}\left[\frac{1}{3}{\bf 1}+\frac{2\beta}{15}(\Lambda+\Lambda^T)+
\frac{16\beta^2}{105}(\Lambda\cdot\Lambda^T)\right]
\label{eq:Gsf}
\eeq
Note that at fixed $\beta$ the gyration tensor has the same N dependence as at 
equilibrium (the scaling exponent is $\nu=0.5$).
The orientation of the chain, as given by eq.(\ref{eq:chi}), when defined 
through the gyration tensor is
\beq
cotg(2\chi_g)=\frac{G_{xx}-
G_{yy}}{2G_{xy}}=\frac{60\beta}{105}=\frac{\beta}{1.75}
\eeq
implying that the orientational resistance $m_g=1.75$ and $\beta$-independent.
This is the well known classical result.

Other used quantities in characterizing chains under SF are the deformation 
ratios
as defined in section \ref{sec:phenom}. From eq.(\ref{eq:Gsf}) we get
$\delta G_{xx}=16/35~\beta^2=0.457 \beta^2; \delta G_{yy}=\delta G_{zz}=0$ as is 
well known.
Note that the above values of the deformation ratios in the laboratory reference 
frame 
imply that in the molecular reference frame both elongation and compression 
directions 
present deformations as follows
\beqa
\delta G_{I}&=&\frac{2}{35}\left[4\beta^2 + 7\beta
\sqrt{1+\left(\frac{4}{7}\beta\right)^2}\right]\\
\delta G_{II}&=& \frac{2}{35}\left[4\beta^2-7\beta
\sqrt{1+\left(\frac{4}{7}\beta\right)^2}\right]
\eeqa
with the following limiting behaviours
\beqa
\delta G_{I}&\rightarrow&\frac{16}{35}\beta^2 \hskip 3.5cm \beta>>\frac{7}{4} \\
\delta G_{II}&\rightarrow&-\frac{49}{140}=-0.35 \hskip 2cm \beta>>\frac{7}{4} \\
\delta G_{I}&=&-\delta G_{II}\rightarrow\frac{14}{35}\beta \hskip 2cm 
\beta<<\frac{7}{4}
\eeqa

\begin{figure}[H]
\centerline{\psfig{file=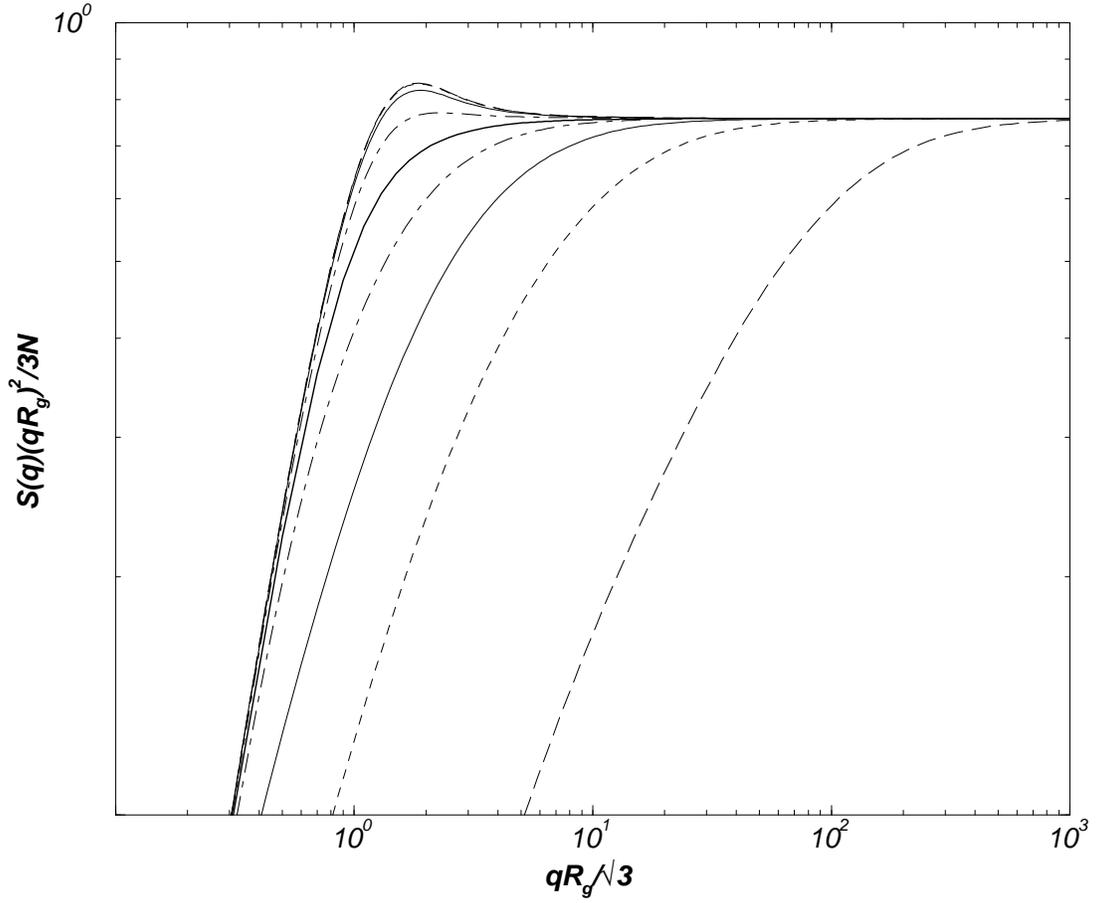,width=12cm,angle=-90}}
\caption{Structure factor of the continuum Rouse chain model under shear at 
various values of 
$\beta$ and for the scattering vectors along the principal axes of the gyration 
tensor in the flow 
plane. $\beta=0.0$, thick line, $\beta=1.0$ dot-dashed lines, $\beta=3.2$ thin 
line, $\beta=10.0$
dashed, and $\beta=100.0$ long dashed. Curves below the equilibrium one are in 
the elongation 
direction I, while curves above the equilibrium one are in the compression 
direction II.}
\label{fig:kratkyrouse}
\end{figure}

Now we go back to our original task, that is the calculation of the chain 
structure factor
under shear flow. As already stated, the double integral over monomers in 
eq.(\ref{eq:sqsf}) with eq.(\ref{eq:ftensor}) cannot be performed analytically 
because 
of the exponential character of the integrand (to be compared with the double 
integral
of the exponent which can be solved and provides the gyration tensor as in 
eq.(\ref{eq:Gsf})).
We applied standard Romberg's method \cite{NumRecF} to numerically evaluate 
those 
integrals in a wide range of $q$'s and for given $\beta$. In figure 
\ref{fig:kratkyrouse}
we report the chain structure factor in its universal form 
($S(\qvec)(qR_g)^2/3N$ 
{\it vs} 
$qR_g/\sqrt{3}$), for various values of $\beta$ and for $\hat\qvec$ along the 
elongation and 
the compression directions in the flow plane (molecular reference frame). We 
note as
in the elongation direction the crossover to the high $q$'s equilibrium power 
law behavior 
strongly depends on the shear rate. The Kratky plot in the compression direction 
presents a $\beta$ dependent overshoot due to the compression of the 
corresponding 
gyration tensor component. This effects however saturates very soon (the curve 
for 
$\beta=3.2$ is already very close to the behavior for $\beta=10$ which is 
indistinguishable
from the one at $\beta=100$) and the crossover to the high $q$'s power law 
behavior
is in practice $\beta$ independent.

In comparison with the stretched ideal chain \cite{PAR97} we note the absence of 
oscillations
in the elongation direction of the sheared chain signal. Such oscillations are 
characteristics of an 
aligned rod with uniform density of scattering centers. 
Our present results show definitively that, even at very high
shear rates, the sheared chain cannot be
thought in terms of such simplified model at variance with Onuki scheme 
\cite{On85}.

Another important characteristics of the structure factor of the stretched 
chain, related to
the presence of tensile blobs, is the Lorentzian shape of the scattering 
function at intermediate
$q$'s as discussed by Benoit {\it et al.}\cite{BDODCFJ75}. In this case one can 
easily show that, 
for large $q$, the scattering function for ${\bf \hat{q}}$ along the stretching 
direction follows the 
equilibrium Ornstein-Zernike behavior $S(q)=1/(1+q^2R_g^2/2)$ if plotted in 
terms of the 
effective wave vector $\tilde q^2=q^2+4/\xi_T^2$, where $\xi_T$ is the tensile 
blob size.
This led Pincus to formulate his scaling for stretched EV chains just changing 
the scaling 
exponent from the ideal to the EV value\cite{P76,PAR97}. Under shear flow the 
chain 
structure factor at intermediate $q$'s in the elongation direction has not a 
Lorentzian 
shape, but rather presents a behavior
$S(x)/N=1/(1+x^2/2+a(\beta)x^{\alpha(\beta)})$ for $x=qR_g/\sqrt(3)\geq 1$, in 
analogy with the 
scattering signal from near critical fluids under shear\cite{On97}.
The fitting function $a(\beta)$ goes linearly from $0$ to $73$ in the interval 
$0\leq \beta \leq 100$,
while $\alpha(\beta)$ goes almost linearly in $log(\beta)$ from $0.62$ to $0.88$ 
in the interval
$3.2\leq \beta\leq 100$. Whether this form for $S(q_I)$ has some theoretical 
foundation 
remains unclear to us.

\section{Excluded volume chain model and Brownian Dynamics 
algorithm}\label{sec:bd}

The ideal chain model of previous section is only the starting point for a more 
realistic description of polymeric chains. When the aim is the study of dilute 
solutions at least two additional effects need to be taken into account namely 
the excluded volume and the hydrodynamic interactions between monomers. In this 
section we describe the model we have used to represent the excluded volume 
effect.
In our previous work on stretched chains \cite{PAR97} we have used a model of 
linear chains with rigid bonds and no angle hindrance. The excluded volume 
interaction where modeled by hard sphere interactions. The diameter $\sigma$ of 
each monomer was chosen as $\sigma=0.65$ in units of the bond length. This model 
exhibits good solvent scaling\cite{PAR97} and is very suitable for Monte Carlo 
calculations. Here we deal with non equilibrium systems for which the stationary 
distribution under shear is unknown and therefore we cannot apply the Monte 
Carlo sampling procedure.
The only way to perform the calculation is by a dynamical technique such as 
Brownian Dynamics(BD). For this technique, the model of the previous work is not 
very appropriate because of the strong discontinuity in the interactions. We 
thus replaced the hard sphere by a Lennard-Jones potential truncated at the 
minimum and shifted in such a way to have a continuous potential
\beqa
v_{EV}(r)&=&v_{LJ}(r)-v_{LJ}(r_c) \hskip 2cm r\leq r_c=2^{1/6}\sigma \\
         &=& 0                    \hskip 4.5cm r>r_c
\eeqa
where $v_{LJ}(r)=4\epsilon[(\sigma/r)^{12}-(\sigma/r)^{6}]$. In our calculation 
$\epsilon$ was the energy unit and $\sigma$ was chosen to be $0.65$ as in the 
previous work.
Moreover the imposition of the bond constraints in BD requires an iterative 
technique such as SHAKE\cite{CR86} that for long chains is quite demanding in 
terms of computer time. We therefore prefer to use flexible rather then rigid 
bonds and represent the bond interaction by an harmonic potential with force 
constant $\chi$ and minimum position $d$: $v_{bond}(r)=\chi~(|\rvec|-d)^2/2$. 
Lennard-Jones interactions between
nearest neighbors along the chain were not considered.

The equation of motion of the i-th monomer is
\beq
\xi \dot{\bf R}_i={\bf F}_i + {\bf f}_i +\xi \kvec\cdot{\bf R}_i
\eeq
where ${\bf F}_i$ is the total force on monomer $i$ deriving from the potential 
energy, and the properties of the noise 
${\bf f}_i$ are 
\beqa
<{\bf f}_i(t)>&=&0 \\
<{\bf f}_n(t){\bf f}_j(t')>&=& 2 k_BT \delta(t-t') \delta_{ij} {\bf 1} 
\label{eq:noisefd}
\eeqa
The above equations can be integrated numerically by the simple finite 
differences scheme
\beq
{\bf R}_i(t+h)=[{\bf 1}+h\kvec]\cdot{\bf R}_i(t) + \frac{h}{\xi}{\bf F}_i(t) + 
{\bf w}_i(t)
+O(h^{3/2})
\label{eq:dlang}
\eeq
where the moments of the white noise ${\bf w}_i$ are 
\beqa
<{\bf w}_i(t)>&=&0 \\
<{\bf w}_i(t){\bf w}_j(t')>&=& 2 h D_0 \delta(t-t') \delta_{ij} {\bf 1} 
\label{eq:noisew}
\eeqa
and $D_0=k_BT/\xi$ is the single monomer diffusion coefficient.
Higher order schemes has been proposed in the literature
which allow to take a larger time step but require more operations per time 
step.
A further advantage of the simple first order scheme over higher order ones, is 
the 
possibility to use a uniformly rather than normally distributed noise, provided
that the zero and second moment are correctly chosen. Indeed, it is easy to 
show that the error introduced by such a substitution is of order $h^{3/2}$ or
higher, that is beyond the precision of the scheme itself.
Use of uniformly distributed noise rather than a gaussian noise saves about 
$50\%$
of the computer time. 
In absence of a systematic comparison on the overall efficiency of integration 
schemes
we have preferred to use the simple one above.

In all our calculations we have fixed $k_BT=1$ and $\xi=1$ which lead to 
$D_0=1$.
The bond force constant has been chosen to be $\chi=40$ and in order to obtain a 
unitary average bond length we had to fix $d=0.945$.

The presence of stiff springs, ensuring the rigidity of the bonds, and strongly 
repulsive core interactions, ensuring the self avoidance and avoiding the self 
crossing of the chain, impose a quite small time step for the stability of the 
numerical scheme.
On the other hand the long relaxation time of polymers and its rapid increase 
with 
the number of monomers ($\tau_N\sim N^{2.2}$ for excluded volume chains without
HI) would require
a time step as large as possible. 
This is the main limitation of any dynamical scheme
when dealing with polymers. In the present work we have chosen a time step
$h=0.00025$ in reduced units. We have tested in some cases that halving the 
time step does not change the results for the chain structure.
Even with such small time step, numerical instabilities occasionally show up.
These occur when two particles end a time step at a reciprocal distance 
considerably
smaller then $\sigma$. The following time step is then driven by very large 
repulsive forces which stretch the bonds related to the pair of particle at a 
distance 
considerably larger then the minimum of the bond potential and so forth.
To cure this occasional pathology we checked at the end of any time step the 
minimum distance between any pair of particles and if it is found to be smaller 
then 
$0.43\sigma$ we reject the time step and we proceed the time integration for a 
time
interval $h$ in many smaller time steps. When the time interval $h$ is reached 
the original time step is restored. With the chosen value for the original time 
step 
we have found that such pathological events happen on average any $2x10^5$ time 
steps 
and that dividing the time step by ten is always enough to escape from the 
instability.  This is a craft made version of more sophisticated adaptive time 
step integration schemes which are however more demanding in terms of computer 
time.

To further minimize the computer time per step we have used a Verlet neighbors 
list \cite{AT86} which is very effective with such short range interactions.

\section{Simulation results}\label{sec:results}

\subsection{Equilibrium scaling of the model}

\begin{figure}[H]
\centerline{\psfig{file=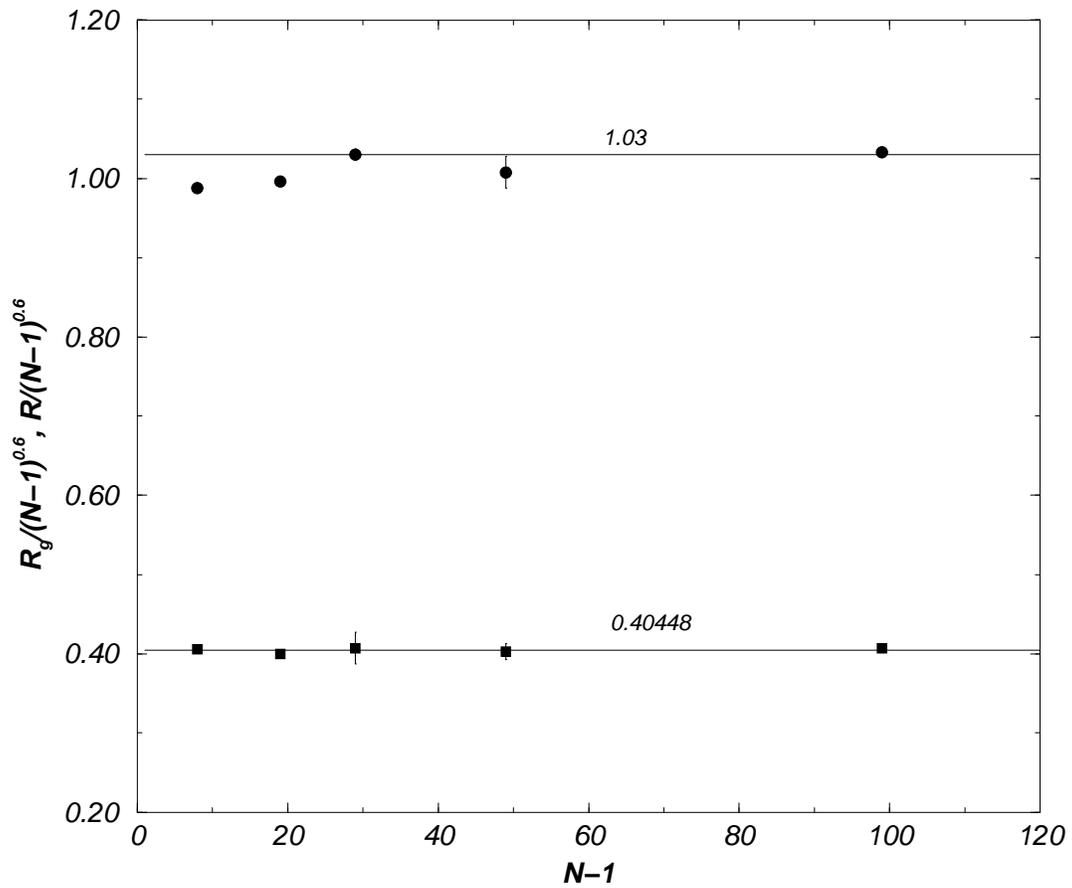,width=12cm,angle=-90}}
\caption{Equilibrium static scaling for the end-to-end distance (circles) 
and the gyration radius (squares).}
\label{fig:eqstatscaling}
\end{figure}

\begin{figure}[H]
\centerline{\psfig{file=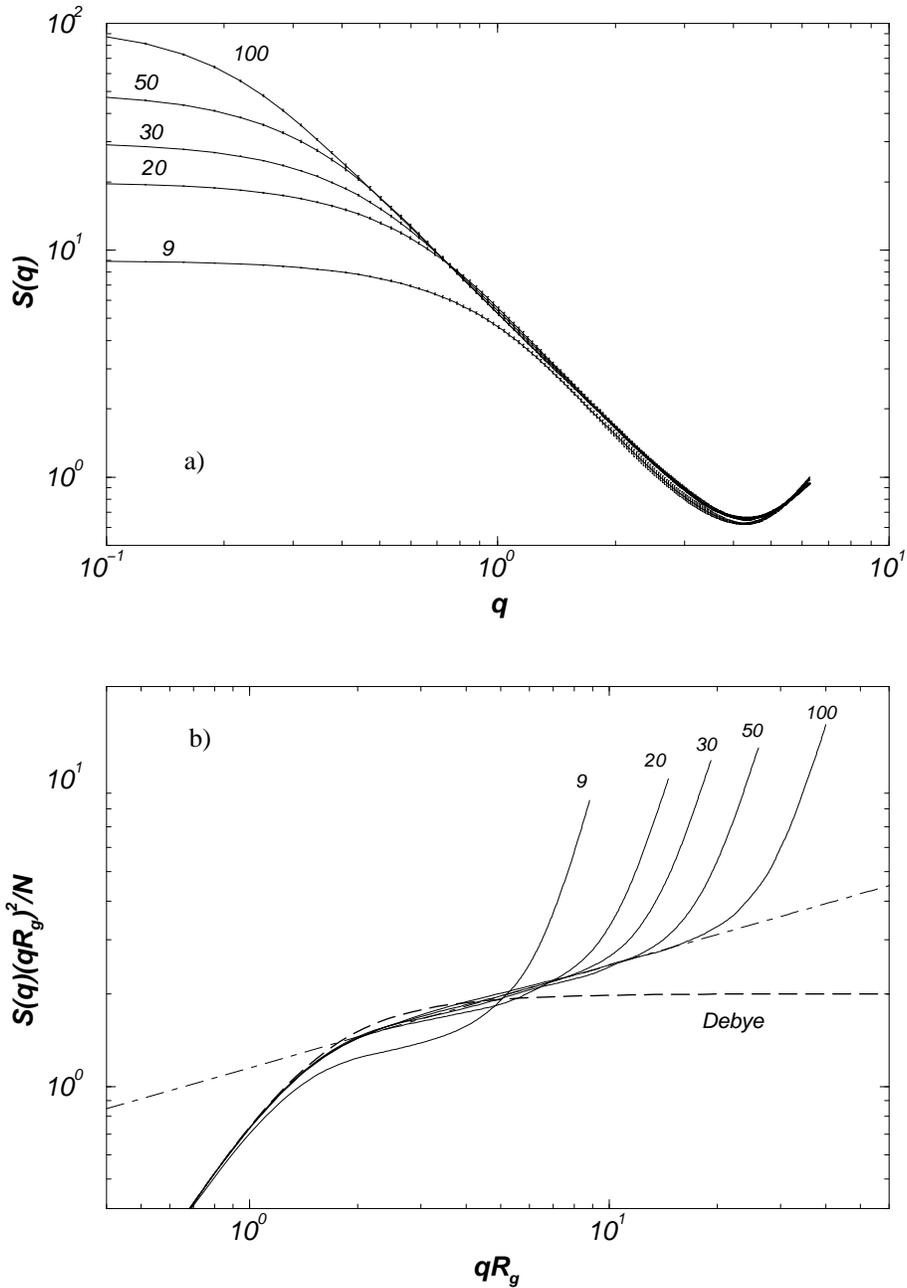,width=12cm,angle=0}}
\caption{a) Static structure factor at equilibrium for various chain lengths 
from $N=9$ to $N=100$.
b) Kratky plot $S(x) x^2/N$ in terms of $x=qRg$. The tick dashed curve is the 
ideal (Debye) behaviour.
The dot-dashed straight line represents the $x^{1/3}$ behavior expected for EV 
chains.}
\label{fig:eqsofq}
\end{figure}

Before proceeding to study the chain model under shear flow we need to 
characterize its equilibrium behavior. In particular we need to know at which 
length the chains start following the known static and dynamic scalings and we 
need to determine the prefactor
in the scaling laws in order to fix the values of the bare shear rate for a
given chain length to work at fixed $\beta$.
In figure \ref{fig:eqstatscaling} we report the radius of gyration $R_g$ and the 
end-to-end
distance $R$ versus the number of links in the chain $(N-1)$. The observed 
scalings are
$R_g=0.40448 (N-1)^{0.6}$, and $R= 1.03 (N-1)^{0.6}$ if we exclude $N=9, 20$. 
Equilibrium 
structure factors for the various chain lengths are shown in figure 
\ref{fig:eqsofq}a.
In figure \ref{fig:eqsofq}b we report the universal plot $(qR_g)^2S(q)/N$ versus
$(qR_g)$. We also add the Debye curve for the ideal chain. We observe that, for 
$N\geq30$, 
the EV chains follow indeed a universal $q^{1/3}$ power law
as predicted on the basis of the static scaling and dimensional arguments 
\cite{DE86}.

\begin{figure}[H]
\centerline{\psfig{file=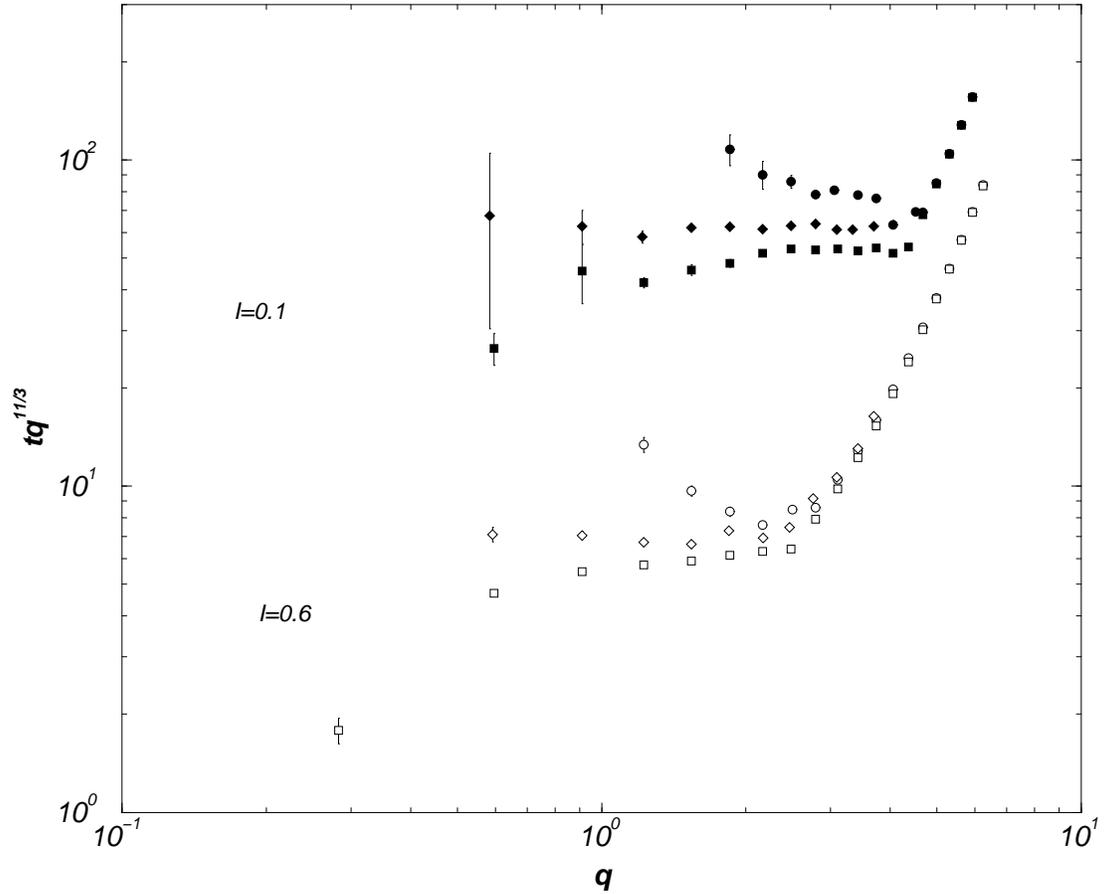,width=12cm,angle=-90}}
\caption{Scaling of the internal dynamics provided by the dynamics structure 
factor as explained in the 
text.$N=20$ circles; $N=30$ squares; $N=100$ diamonds. Closed symbols are for 
$I=0.1$ while open symbols
corresponds to $I=0.6$.}
\label{fig:eqdynscaling}
\end{figure}

By construction, the chain centre of mass diffusion coefficient is 
$D_{cm}=D_0/N\propto N^{\nu'}$ where $\nu'$ is the dynamic exponent, which
provides a characteristic time of the chain 
$\tau_N=R_g^2/(6 D_{cm})=(0.40448)^2 (N-1)^{1.2}N/6D_0\sim N^{2\nu+1}$. Here 
$\nu=3/5$ is the 
Flory scaling exponent for EV chains, and $\nu'=1$ as HI are absent. 
In order to check the consistency between global and internal dynamics we 
extracted the 
characteristic time of the latter by inverting the chain dynamic structure 
factor 
$I(q,t)=S(q,t)/S(q)$ as we have previously done for MD chains 
\cite{PR91-92,RP99}. 
Indeed scaling and dimensional arguments predict that $I(t,q,N)$ be 
a universal function of $(tq^{-x})$ with $x=2+\nu'/\nu=11/3$ in the present 
case.
Figure \ref{fig:eqdynscaling} shows the behavior of $(tq^{11/3})$ vs $q$ 
for two values of 
$I$ and for various chain lengths. We observe that for $N=20$ the chain does not 
exhibit the correct scaling of internal times while from $N=30$ on, both static 
and dynamics are in the scaling laws regime. The data we show are results of 
quite long
runs as reported in table \ref{table1}.

\subsection{Chain structure at fixed $\beta$}\label{sec:NEBD}

In this subsection we describe the results for the chain structure under shear 
flow for various chain lengths at the same value of the reduced shear rate 
$\beta$ that we defined 
as $\beta=\dot\gamma R_g^2/(6 D_{cm})$. We have limited our study to two values 
of
$\beta$, namely $\beta=3.2$ as in our previous MD study and in several 
experiments\cite{LO88,LM99},
and $\beta=10$. In tables \ref{table2} and \ref{table3}, we report various 
technical details 
of our simulations
and the results for the orientation and the deformation.
We studied chains up to $N=300$ which require very long runs (up to $3~10^9$ 
time steps for the 
EV case, see table \ref{table3}).

\begin{figure}[H]
\centerline{\psfig{file=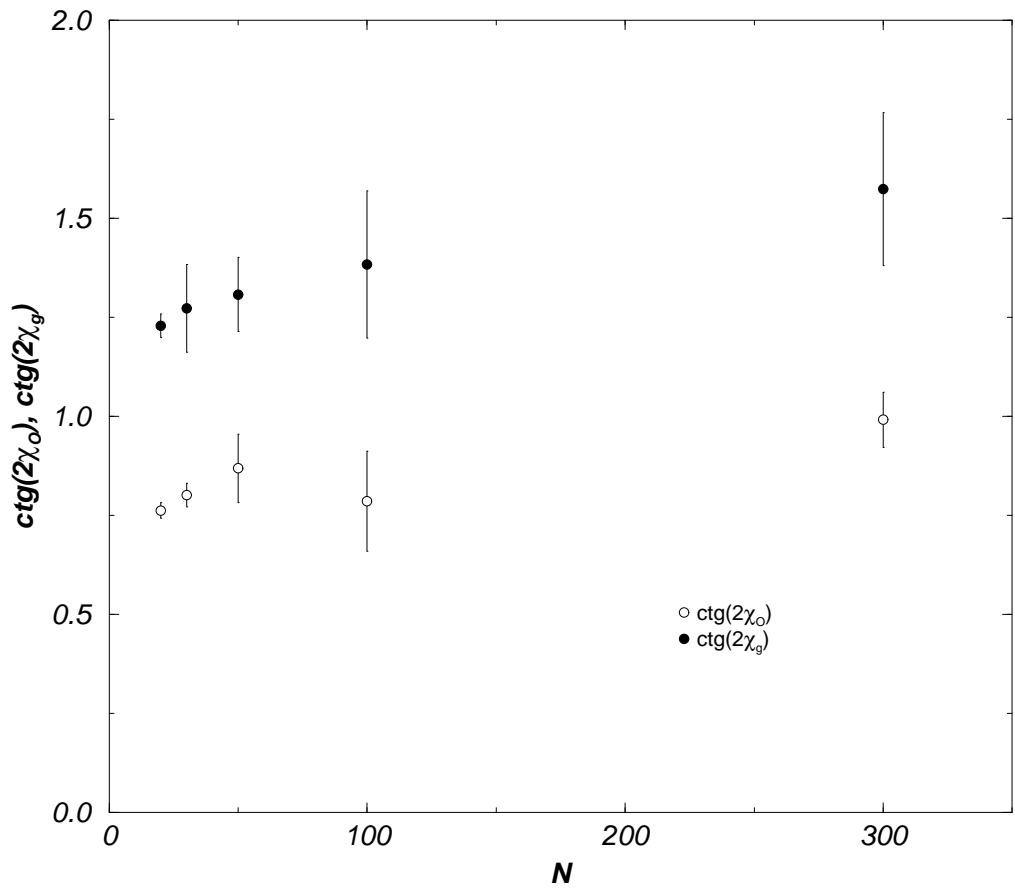,width=12cm,angle=-90}}
\caption{Orientation angle (closed symbols) and extinction angle (open symbols) 
at $\beta=3.2$ versus $N$.}
\label{fig:b3.2orient}
\end{figure}

As in our previous MD study of short chains (up to $N=50$), we analyze the 
results
in the molecular reference frame on account of the observation that the 
orientational angle depends
on $N$ and $\dot\gamma$ through $\beta$ only. This is also suggested by the 
Rouse model results 
of section \ref{sec:rouse} and by its extension when HI at the level 
of equilibrium
preaveraging are considered \cite{Carl}.
Our new results for EV chains at $\beta=3.2$ confirm this expectation within 
error bars, although short chains seems to be slightly less oriented as shown in 
figure \ref{fig:b3.2orient}. 
On the same figure we report the extinction angle related to the birefringence 
which also is found to be the same within error bars for all $N$ at fixed 
$\beta$.

Before discussing the results for our chain model with EV interactions
it is instructive to consider the same coarse grained model with
the EV interactions switched off ($\epsilon=0$ in the LJ potential).
Comparison of the chain structure factor of such a model with the 
continuous ideal model result eq.(\ref{eq:sqsf}) (not shown here) 
validates the theory itself and 
the numerical procedure applied to compute the double integral.
Moreover data for the orientation and the deformation of this model 
can be useful in understanding qualitatively the role of the 
finite extensibility effects due 
to the stiff harmonic potential which keeps the bond length nearly constant.  
Data for the orientation and deformations of ideal chains with $N=100$ and 
$N=300$ at 
$\beta=3.2$ and $\beta=10$ are reported 
in table \ref{table2} and 
compared to the Rouse model predictions. At $\beta=3.2$, 
the orientation for all chain lengths is in agreement with the 
long chain limit while the deformations for $N=100$ deviate considerably from 
the ones 
for $N=300$ (except in the direction II)
which are instead in agreement with the Rouse prediction. 
Interestingly, the main eigenvalue $G_I$ of the 
gyration tensor follows an apparent power law $N^{1.5}$ qualitatively similar to 
the behavior
observed for MD data on shorter chains with EV and HI\cite{PR95}.
At $\beta=10$ both chains exhibit finite extensibility effects even on the 
orientation.

Now we turn to the EV results. In table \ref{table3} we collect the data for the 
orientation and the deformation of our chains. We also report data for the 
longest chains studied in ref. \cite{PR95} by molecular dynamics. 
We note at $\beta=3.2$, that the 300-bead chain appears to be 
slightly more oriented and deformed than the shorter chains, although the noise 
is quite large. At $\beta=10$ instead, the 300-bead chain is considerably more 
oriented and deformed than the 100-bead chain, showing  an important influence 
of the finite extensibility. Comparison with previous MD data for N=50 shows a 
general agreement for all quantities.

From these results and those of table \ref{table2}, it appears reasonable to 
assume that the N=300 results at $\beta=3.2$ are representative of the global 
properties of long chains, namely that the remaining small systematic finite 
chain effects are masked by statistical errors of a few percents (due to finite 
statistics). Comparing EV chains to $\theta$ chains at the same reduced shear 
rate, 
we observe for EV chains a less marked alignment with respect to the flow 
(larger $\chi$ value) and a smaller global deformation, in qualitative agreement 
with recent LS results \cite{LM99}.

\begin{figure}[H]
\centerline{\psfig{file=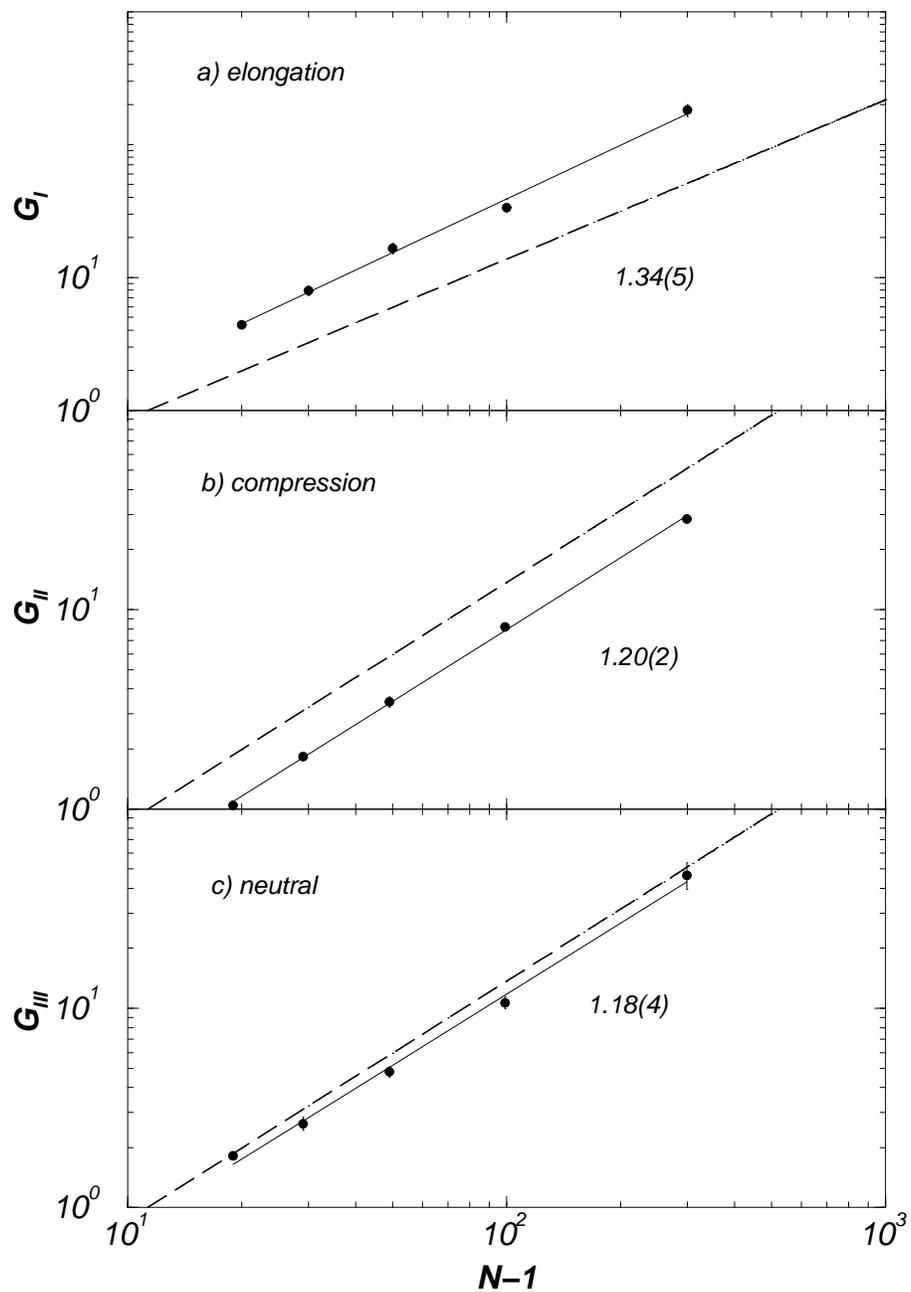,width=12cm,angle=0}}
\caption{Scaling of the eigenvalues of the gyration tensor in the molecular 
reference frame. 
The equilibrium behavior is represented as a thick dashed line. 
Continuous lines are power law fits to the data.
The apparent scaling exponents in the three directions are reported on the 
plots}
\label{fig:b3.2GvsN}
\end{figure}

An important question concerns the universal character of the structure factor 
under 
shear flow. It is well known that at equilibrium the quantity $S(q,N)/N$ is a 
universal function of $qR_g$ only, for any coarse-grained model and for long 
enough chains\cite{DE86}. 
In the case of the chain stretched at both ends, we have shown that different 
universalities in terms of $qR_g$ hold separately in the longitudinal 
and the transverse directions\cite{PAR97}. Alternatively one could investigate 
self-similarities in terms of $q\sqrt{G_{\alpha}}$ where $G_{\alpha}$ are the 
diagonal components of the gyration tensor. For the chain stretched by its ends, 
it is well known that fluctuations of the end-to-end distance (and therefore 
$G_{\alpha}$) at fixed reduced force scale with $N$ as the equilibrium radius of 
gyration and therefore the above two alternatives are equivalent. 
In our previous study of linear chains under shear flow we observed apparent 
scaling exponents $\nu_{\alpha}$ for the eigenvalues of the gyration tensor with 
$(N-1)$ which were different from the equilibrium one in the flow plane 
directions while in the neutral (out of plane) direction the equilibrium scaling 
was preserved within error bars\cite{PR93,PR95}.
In the present study on longer chains we find that at $\beta=3.2$ the 
equilibrium scaling $G_{\alpha}\propto N^{2\nu}$ is preserved
in the neutral direction ($\alpha=III$) and in the in-plane compression 
direction ($\alpha=II$) (see figure \ref{fig:b3.2GvsN}).
In the elongation direction an apparent scaling exponent higher than the 
equilibrium value is still 
observed ($\nu_I\sim0.7$) although the noise is considerably higher than in the 
other two principal directions. The results for the chain without EV 
interactions mentioned above demonstrate that such apparent scaling is related 
to finite extensibility effects rather than to EV interactions. For this reason 
we investigate universality in terms of $qR_g$.

\begin{figure}[H]
\centerline{\psfig{file=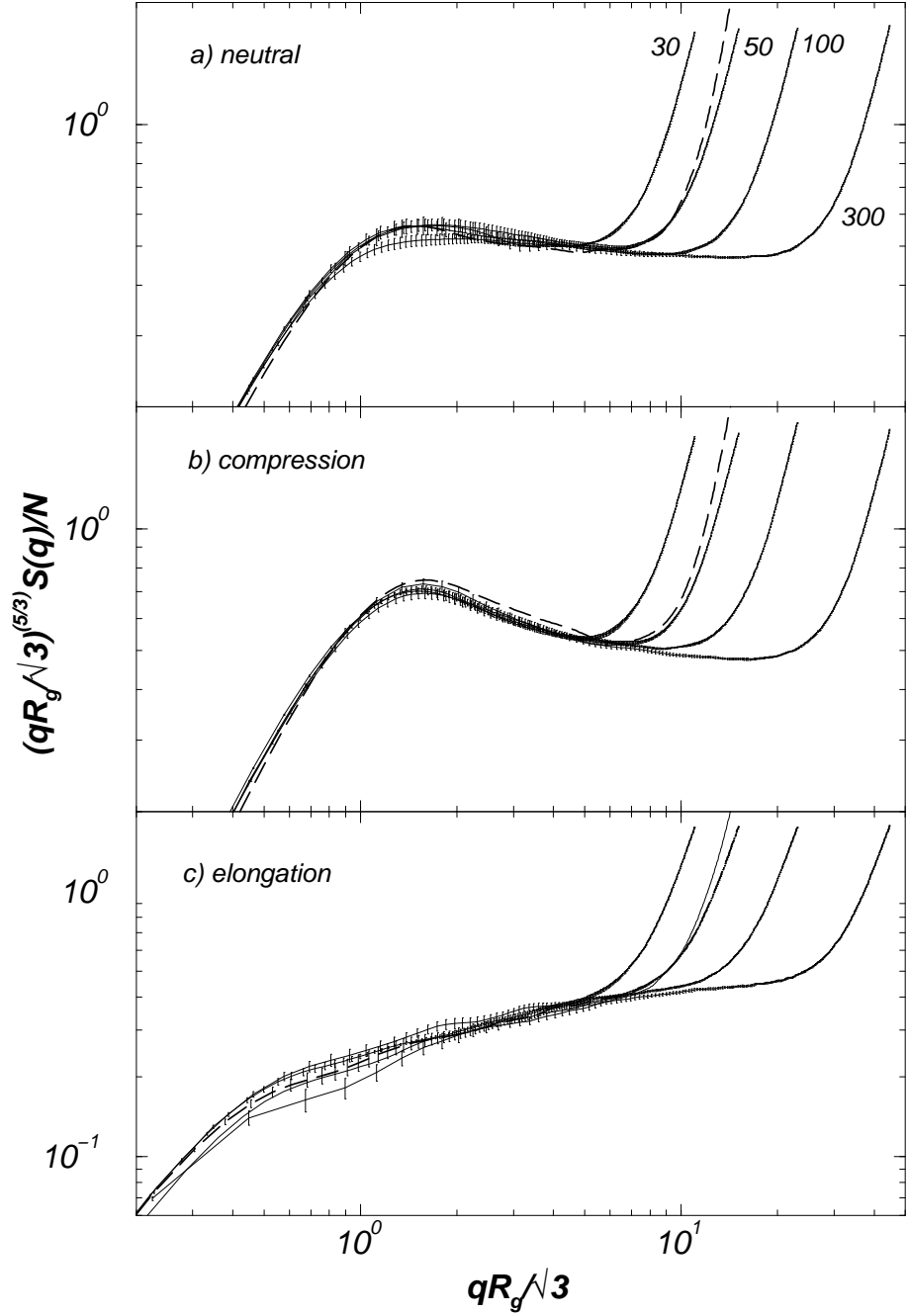,width=12cm,angle=0}}
\caption{Structure factor for the EV model at $\beta=3.2$ and
for various $N$ as indicated in the figure. MD data from ref.[38] for 
$N=50$ at $\beta=3.2$ 
are reported as a thick dashed curve.}
\label{fig:b3.2univeveq}
\end{figure}

In figure \ref{fig:b3.2univeveq}a,b,c 
we plot, at $\beta=3.2$ and for $30\leq N\leq 300$, 
$(qR_g/\sqrt{3})^{5/3}S(q)/N$ versus 
$qR_g/\sqrt{3}$ for $\alpha=III,II,I$ respectively and we compare with  
the MD data for N=50 at the same $\beta$ from ref.\cite{PR95}. 
For the latter we have used the observed 
equilibrium exponent $\nu_{MD}=0.57$ rather then the Flory classical value 
$\nu_F=0.6$.
Note that in that model the solvent was explicitly considered so that deviations 
from the 
mean field value are not contradictory. 
We observe that, with the exception of $N=30$ in the neutral direction, 
universality is indeed observed within error bars in all directions.
A general feature of those curves is the presence of a quite slow crossover back 
to the equilibrium behavior at large $q$'s which would be represented by an 
horizontal line. 
That $N=300$ is still too short to cover all length scales of the problem is 
clearly 
shown by the fact that even for this length the equilibrium behavior is not 
completely recovered before the chain finite extensibility shows up, a sign that 
finite chain length effects are still present. While in the neutral direction 
the effect of the flow is tiny,
in the in-plane directions
the flow strongly changes the scattering function. 
We note that the noise
level is considerable higher in the elongation direction than in the other 
directions.
Moreover statistical convergence of the results at low $\beta$ appears to be 
considerably slower
than at high $\beta$ and at equilibrium (see also table \ref{table3}). This is 
compatible with
the possible appearance of new characteristic relaxation times in weak flow as 
indicated
by recent experiments \cite{LHBW99}.

Finally we note that the structure factor obtained in our previous MD simulation 
on a different chain model embedded in an explicit atomic solvent under shear 
flow \cite{PR95}, follows very closely in all directions the universal behavior 
predicted by the present BD simulations. As the former study includes 
automatically the hydrodynamics interactions \cite{PR91-92,DK91-93} while the 
present
calculation do neglect them from the start, it appears that HI play a secondary 
role on the chain structure under shear flow, once one works at a fixed value of 
the suitably defined reduced shear rate.
Because of the limited length of the largest chain studied by molecular 
dynamics (N=50), we could not detect on the chain structure factors the slow 
trend back 
to the equilibrium behavior at large q's, which explains our 
previous 
analysis in terms of anisotropic scaling laws.

\begin{figure}[H]
\centerline{\psfig{file=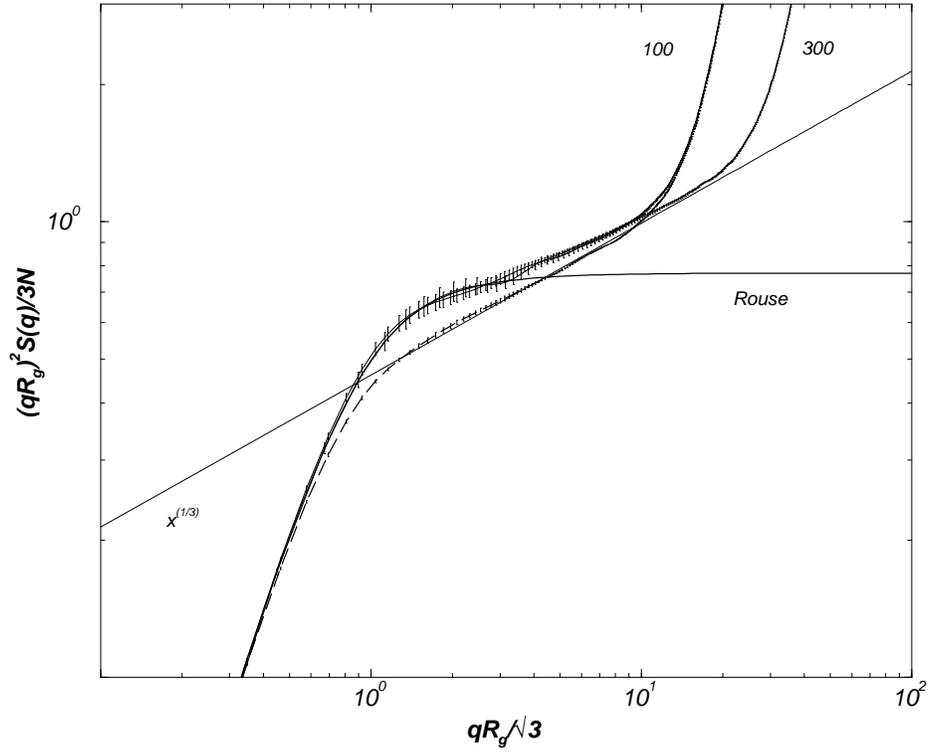,width=10cm,angle=-90}}
\caption{Ideal Kratky plot in the neutral (out of plane) direction III at 
$\beta=3.2$
for $N=100$ and $N=300$. The thick dashed curve is the equilibrium behavior 
for $N=100$. The Rouse behavior and the asymptotic EV scaling are 
represented as continuous lines.}
\label{fig:b3.2univideqIII}
\end{figure}

\begin{figure}[H]
\centerline{\psfig{file=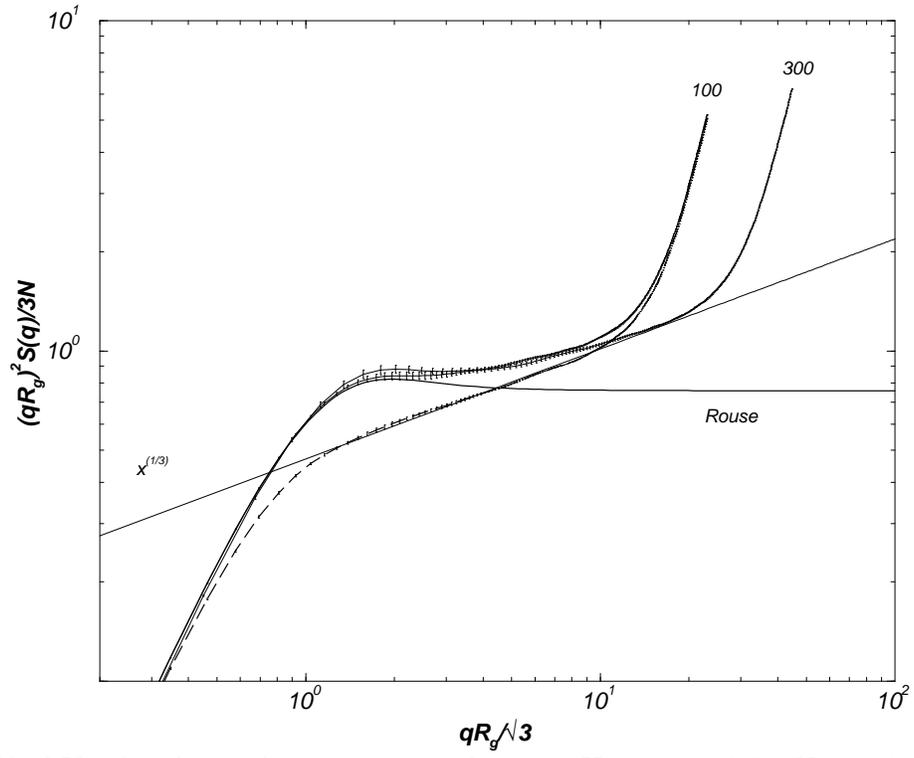,width=10cm,angle=-90}}
\caption{Ideal Kratky plot in the compression direction II at $\beta=3.2$ for 
$N=100$ and $N=300$. 
Symbols are like in figure \ref{fig:b3.2univideqIII}.}
\label{fig:b3.2univideqII}
\end{figure}

\begin{figure}[H]
\centerline{\psfig{file=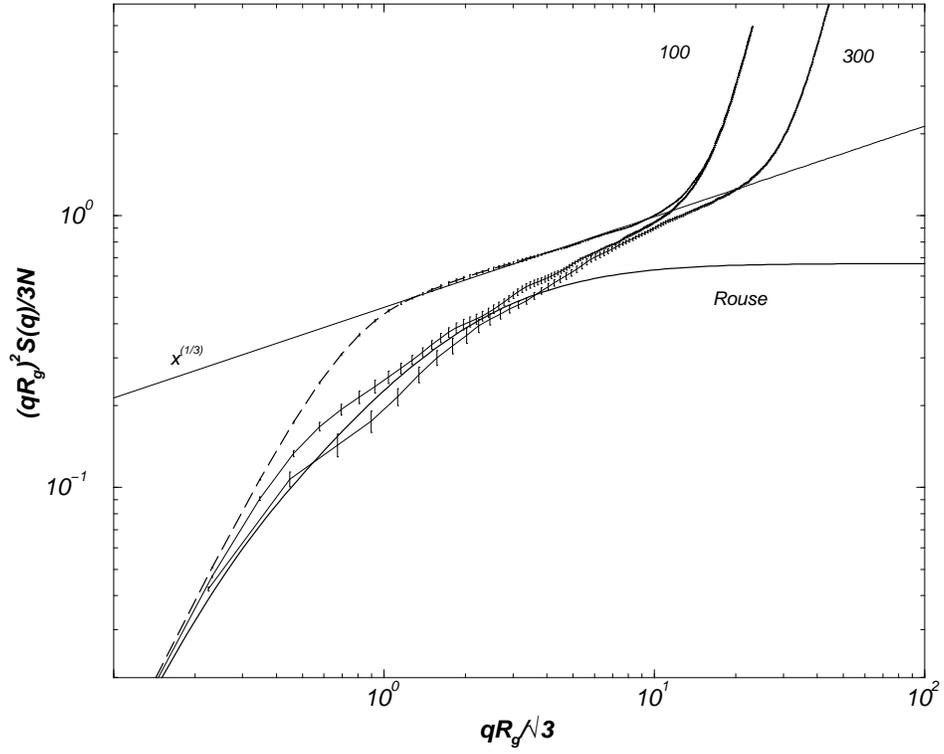,width=10cm,angle=-90}}
\caption{Ideal Kratky plot in the elongation direction I at $\beta=3.2$ for 
$N=100$ and $N=300$. 
Symbols are like in figure \ref{fig:b3.2univideqIII}.}
\label{fig:b3.2univideqI}
\end{figure}

In order to test the ``shear'' blob hypothesis we report in figures 
\ref{fig:b3.2univideqIII},\ref{fig:b3.2univideqII},\ref{fig:b3.2univideqI},
$(qR_g)^2S(q)/3N$ versus $qR_g/\sqrt{3}$ for $N=100$ and $N=300$ at $\beta=3.2$ 
and 
for $N=100$ at equilibrium, 
and we compare with the ideal chain behavior. 
In all directions Rouse behavior describes the excluded volume chain up to 
$qR_g/\sqrt{3} \sim 3$. Similarly to the stretched 
chain case \cite{PAR97}, we clearly see in the compression and the neutral 
directions   
a crossover from ideal statistics at low $q$'s to EV statistics at high $q$'s.
The inset of EV statistics is at $qR_g/\sqrt{3} \sim 9$ in the two directions.
Note that for $N=100$ the EV regime is completely missing and even with $N=300$ 
it is very narrow 
in the compression direction. In the elongation direction the noise level is 
quite high in the 
Rouse (small $q$) regime. As stated above, in this direction flow effects 
extends to higher $q$'s 
than in the other directions, above $qR_g/\sqrt{3}=9$. On the other hand, the 
finite extensibility 
takes place around $qR_g/\sqrt{3}=20$ for $N=300$ and therefore the EV regime 
cannot be observed 
in this direction for such chain length. 

\begin{figure}[H]
\centerline{\psfig{file=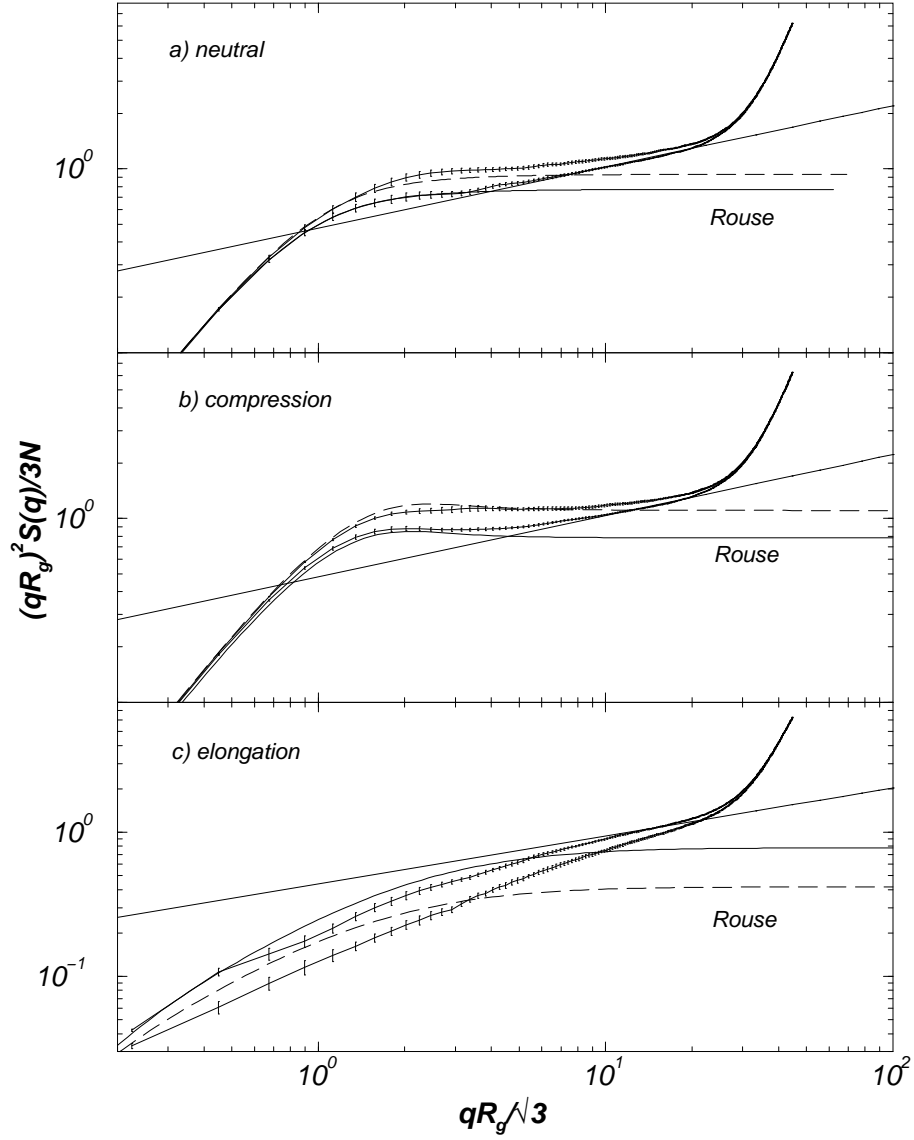,width=12cm,angle=0}}
\caption{Ideal Kratky plot in the neutral direction III for $N=300$ at 
$\beta=3.2$ (lower curve) 
and $\beta=10$ (upper curve). The ideal behaviours, as well as the asymptotic EV 
scaling are 
reported.}
\label{fig:n300univid}
\end{figure}

Onuki's phenomenological model \cite{On85} is based on the assumption that below 
a 
characteristic length scale $\xi$ the effects of the flow are negligible. The 
length scale separation
is dictated by the typical relaxation time of chain density fluctuations at that 
scale.
At a given shear rate $\dot\gamma$, the crossover occurs at a scale $\xi=A 
n_c^{\nu}$ such that 
the longest relaxation time of the blob is $\tau_c=\xi^2/(6D_c)=\dot\gamma^{-1}$ 
where 
$D_c=D_0/n_c^{\nu'}$ is the diffusion of the blob as a whole and $\nu'$ is the 
dynamic scaling exponent.
Working at fixed reduced shear rate $\beta=\tau_N\dot\gamma$, the number of 
monomers per blob
results in $n_c=N/\beta^{1/(2\nu+\nu')}$ so that the number of blobs per chain 
is 
$N_b=N/n_c=\beta^{1/(2\nu+\nu')}$ and 
$\xi=A N^{\nu}/\beta^{\nu/(2\nu+\nu')}=R_g/\beta^{\nu/(2\nu+\nu')}$. 
For our model ($A=0.40448, \nu=3/5, \nu'=1$) at $\beta=3.2$ we have $N_b=1.7, 
n_c=177$ and
$\xi=9.02$ for $N=300$. 
The crossover from ideal to EV statistics in figure \ref{fig:b3.2univideqIII} is 
observed
around $qR_g/\sqrt{3}=4.0$ which for $N=300$ provides $q_c=0.56\sim 5/\xi \sim 
2\pi/\xi$. 
Therefore the shear blob hypothesis is compatible with our results although a
more firm test on the basis of longer chains studied at various values of 
$\beta$ is still missing.
To finally prove that our chains are too short for this purpose we show in 
figure 
\ref{fig:n300univid} the Kratky plots 
in the three principal directions for $N=300$ and $\beta=3.2$ and $\beta=10.0$.
It is evident that at $\beta=10.0$ the finite extensibility effect interferes 
with  
the EV regime giving rise to anomalous behaviours. 
 
\section{Conclusions}\label{sec:conclusions}

In order to study the structure of a single chain subjected to a homogeneous and 
steady shear flow and to elucidate the effects of excluded volume interactions 
we have computed the scattering function of the continuous Rouse chain model in 
flow and we have performed BD simulations of a chain model of Fraenkel springs 
with fully developed EV interactions.

Concerning global properties, our simulations confirm that at a given reduced 
shear rate, EV chains are less deformed and less oriented along the flow lines 
than chains at the $\Theta$ point \cite{LM99}.

We have shown that, even in the ideal case, the chain structure in shear flow 
cannot be 
explained by the known behavior of a chain stretched at its ends. In particular 
we have shown that the structure factor of the Rouse model under SF for the 
scattering vector lying along the in plane elongation direction is very 
different from the one of the Rouse chain stretched at its ends with $q$ along 
the stretching direction. It does not present the oscillations typical of 
permanent extension, but it is a monotonous curve qualitatively similar to the 
equilibrium Debye curve (see figure \ref{fig:b3.2univideqI}). In the elongation 
direction, the onset of the well known $q^{-2}$ behavior signaling the static 
scaling of ideal chains is shifted to higher $q$'s than at equilibrium and is 
strongly dependent on the shear rate. In the compression direction 
it is at the same location as at equilibrium and it is almost insensible to 
shear rate.

Adding the EV interactions, our results suggest that the universality of 
$S(\qvec,N,\dot\gamma)$ in terms of $x=\qvec R_g$ and $\beta$ only ($R_g$ is the 
equilibrium radius of gyration and 
$\beta$ is computed on the basis of the equilibrium characteristic relaxation 
time), exhibited by the Rouse model at any flow intensity, is preserved by the 
EV interactions provided the chains are long enough. Studying chains between 30 
and 300 monomers we indeed observe that the eigenvalues of the gyration tensor 
in the compression (II) and in the neutral (III) directions follow scaling laws 
with N with the equilibrium Flory exponent.
In the elongation direction a larger apparent scaling exponent is instead 
measured (around 0.7), in agreement with previous MD data\cite{PR95,RP99,AKH99} 
which led us to propose an anisotropic scaling picture for intermediate reduced 
shear rates\cite{PR95}. 
From the results for the same chain model without EV interactions we 
infer that this anomalous scaling arises from finite extensibility effects 
rather than EV interactions so as the equilibrium scaling should be recovered 
even in the elongation direction for longer chains. Unfortunately the dynamical 
character of the BD algorithm and the unfavorable scaling with N of the chain 
relaxation time does not allow us to study chains longer than $N=300$.

The general agreement found between our present data (where HI are absent from 
the model) and the previous non-equilibrium MD data for short chains (where HI 
are automatically included) strongly suggests that HI effects play a minor role 
on the chain structure under 
shear flow, provided we compare results at the same reduced shear rate, which 
incorporates implicitly the HI effect through the longest relaxation time of the 
chain at equilibrium.
The EV and the finite extensibility effects  are by far the most 
relevant effects to be taken into account. 

For our longest chain at $\beta=3.2$, the structure factor in the neutral (III) 
and in the compression (II) directions exhibit a clear crossover from the ideal 
behavior at small $q$'s, well predicted by the Rouse model, to the isotropic 
equilibrium 
EV behavior at high $q$'s, characterized by a power law $S(q)\sim q^{-1/\nu}$ 
with $\nu=\nu_{eq}=3/5$. This behavior is usually related to the presence of 
the so called ``blobs'', in the present case``shear blobs''. Blobs were detected 
in the chain structure factor either experimentally for semi-dilute 
solutions 
\cite{FBC78} or by Monte Carlo simulations for a single EV chain 
stretched at 
its ends \cite{PAR97}. The presence of blobs in a chain under shear 
flow has 
been predicted  theoretically \cite{On85} but had never been 
observed so far, neither by SANS nor by simulation. The results of 
our BD simulations are compatible with the existence of blobs (see 
section \ref{sec:NEBD}) but, because of finite chain size effects, a 
quantitative  test of this hypothesis would require chain lengths 
larger than N=300. It is interesting  to note that the largest chain 
considered  here has a size which corresponds to a polystyrene  
chain of the order of 300 Kuhn segments of about $18 \AA$, 
which yields a molecular mass of 300.000 Daltons \cite{LL00}.

The present analysis should be tested by scattering  
experiments on dilute solutions for chains in good solvent 
subjected to steady shear flow. It would be interesting to study 
at low $qR_g$, either by Light or by Neutron scattering,  
whether the Rouse model indeed describes the global orientation 
and deformation of chains for 
large deformations, i.e. at $\beta>>1$. Remaining 
discrepancies could be interpreted as a direct measure of FE
and HI influence on these structural properties far from the equilibrium 
structure. Ultimately, the most spectacular experimental result we 
seek for remains a SANS study of a suitable ``polymer-good solvent''
pair which would allow us to follow the form factor of a polymer of 
a few million Daltons under shear flow at large $\beta$ over a broad 
$q$ regime. This would allow a direct observation  of the crossover 
from ideal chain to EV chain statistics across a crossover  
diffusion vector $q_c$ characterizing the shear blob dimension in 
a particular direction.

Concerning future simulation work, a more quantitative test of HI effects 
on the chain structure is in order and could be performed by BD on short 
chains.

\section{Acknowledgments}
We thank M. Baus, B. D\"unweg, P. Lindner and R.Winkler for useful discussions 
and suggestions. 

\appendix

\section{Ideal chain under steady shear flow}\label{app:ssd}
In this appendix we derive the steady state distribution function of the 
continuous gaussian 
chain model under an homogenous shear flow, eqs. (\ref{eq:PsiX}), 
(\ref{eq:psipf}) and (\ref{eq:betap}).

The dynamics of the model under a generic flow field, specified by the velocity 
gradient tensor ${\bf k}$, is defined by the Langevin equation\cite{DE86}
\beq
\xi \frac{\partial{\bf R}_n(t)}{\partial t}=
\chi \frac{\partial^2{\bf R}_n(t)}{\partial n^2}
+ {\bf f}_n(t) + \xi {\bf k}(t)\cdot {\bf R}_n(t)
\label{eq:langrouse}
\eeq
with the boundary conditions
\beq
\left[\frac{\partial{\bf R}_n(t)}{\partial n}\right]_{n=0} = 
\left[\frac{\partial{\bf R}_n(t)}{\partial n}\right]_{n=N} = 0
\eeq 
${\bf f}_n(t)$ is the gaussian random force acting on the chain and is defined 
by 
\beqa
<{\bf f}_n(t)>&=&0 \\
<{\bf f}_n(t){\bf f}_m(t')>&=& 2 k_BT \delta(t-t') \delta(n-m) {\bf 1} 
\label{eq:noisef}
\eeqa
where $<...>$ indicates an average over the distribution of the random noise and 
${\bf 1}$ is the unit tensor.
As it is well known the above Langevin dynamics corresponds to the following 
diffusion
equation (Smoluchowsky) for the distribution function in the chain 
configurational 
space\cite{DE86,Z69}
\beq
\frac{\partial\Psi}{\partial t} = -\int_0^N dn 
\nabla_{{\bf R}_n}\cdot \left[\dot{\bf R}_n\Psi\right]
\eeq
with the dynamics of monomers given by
\beq
\dot{\bf R}_n=\frac{\chi}{\xi}\frac{\partial^2{\bf R}_n}{\partial n^2}-
\frac{k_BT}{\xi}\nabla_{{\bf R}_n} [ln\Psi] + {\bf k}\cdot{\bf R}_n
\eeq
Let us introduce now the normal modes defined as
\beq
{\bf X}_p(t)=\frac{1}{N}\int_0^N dn~cos\left(\frac{p\pi n}{N}\right){\bf R}_n(t)
\hskip 3cm p=0,1,2,....
\label{eq:R2X}
\eeq
in terms of which the chain positions are given as
\beq
{\bf R}_n(t)={\bf X}_0(t)+2\sum_{p=1}^\infty cos\left(\frac{p\pi 
n}{N}\right){\bf X}_p(t)
\hskip 3cm 0\leq n\leq N
\label{eq:X2R}
\eeq
The completeness relations for the normal modes basis are
\beqa
\delta_{pq}= \frac{2}{N}\int_0^N dn~cos\left(\frac{p\pi n}{N}\right) 
cos\left(\frac{q\pi n}{N}\right) \\
N \delta(n-m)=1+2\sum_{p=1}^\infty cos\left(\frac{p\pi n}{N}\right)
cos\left(\frac{p\pi m}{N}\right)
\eeqa
Note that, for homogeneous flows which we consider here, the normal modes 
of the dynamics are the same as at equilibrium.

The Langevin equations for the normal modes are
\beq
\xi_p \partial_t {\bf X}_p(t)= -\chi_p {\bf X}_p(t) + \tilde{\bf f}_p(t)
+\xi_p {\bf k}\cdot{\bf X}_p(t)
\eeq
with the following definition of the symbols\cite{DE86}
\beqa
\xi_0 &=& N\xi, \hskip 4cm \xi_p=2N\xi \hskip 2cm p=1,2,3... \\
\chi_p &=& \frac{6\pi^2k_BT}{Nd^2}p^2=\frac{2\pi^2\chi}{N}p^2=\chi_1 p^2\\
<\tilde{\bf f}_p(t)>&=&0, \hskip 1cm 
<\tilde{\bf f}_p(t)\tilde{\bf f}_q(t')>=2\xi_p k_BT \delta_{pq}\delta(t-t'){\bf 
1}
\eeqa

The distribution function of normal modes is factorized
\beq
\hat\Psi(\{{\bf X}\},t)=\prod_{p=0}^\infty \hat\psi_p({\bf X}_p,t)
\eeq
and one can easily derive the following diffusion equation for the $p-th$ mode
($p>0$)
\beq
\partial_t\hat\psi_p({\bf X}_p,t)=-\nabla_{{\bf X}_p}\cdot 
\left[({\bf k}\cdot{\bf X}_p - \frac{\chi_p}{\xi_p}{\bf X}_p)\hat\psi_p -
\frac{k_BT}{2\xi}\nabla_{{\bf X}_p}\hat\psi_p\right]
\label{eq:psip}
\eeq
The center of mass motion is instead described by the diffusion equation for 
the zero mode
\beq
\partial_t\hat\psi_0({\bf X}_0,t)=-\nabla_{{\bf X}_0}\cdot 
\left[{\bf k}\cdot{\bf X}_0 \hat\psi_0 -
\frac{k_BT}{2\xi}\nabla_{{\bf X}_0}\hat\psi_0\right]
\label{eq:psi0}
\eeq
The solution of eqs.(\ref{eq:psip}) is a multivariate guassian 
\beq
\hat\psi_p({\bf X}_p,t)=\left(\frac{1}{(2\pi)^3 det[\alpha_p(t)]}\right)^{1/2}
e^{ -\frac{1}{2}{\bf X}_p\cdot\alpha_p^{-1}(t)\cdot{\bf X}_p }
\eeq
where the time dependent covariance matrix $\alpha_p(t)$ is given by (see 
eq.(7.161) in ref.\cite{DE86})
\beq
\alpha_p(t)=\int_{-\infty}^t dt' \frac{2k_BT}{\xi_p} {\bf E}(t,t')\cdot{\bf 
E}^T(t,t')
e^{-\frac{2p^2}{\tau_1}(t-t')}
\label{eq:alphap}
\eeq
Here $\tau_1=\xi_1/\chi_1$ is the relaxation time of the first normal mode
and the tensor ${\bf E}(t,t')$ is the deformation tensor of the 
external field defined by 
\beq
{\bf E}(t,t')={\bf 1} + \int_{t'}^t dt''{\bf k}(t'')
\eeq
For the steady shear flow to which we are interested in this work we have
\beq
{\bf k} = \left (
\begin{array} {ccc} 0 & \dot\gamma & 0\\
                    0 &       0    & 0\\
                    0 &       0    & 0
\end{array}
\right )
\eeq
\beq
{\bf E}(t,t') = \left(
\begin{array} {ccc} 1 & \dot\gamma(t-t') & 0 \\
                    0 &        1         & 0 \\
                    0 &        0         & 1 
\end{array}
\right)
\eeq
so that the covariance matrices $\alpha_p$ results time independent and takes 
the form
\beq
\alpha_p=\frac{k_BT}{\chi_p}\left[{\bf 1}+\frac{\tau_1}{2p^2}(\kvec+\kvec^T)
+2\left(\frac{\tau_1}{2p^2}\right)^2\kvec\cdot\kvec^T \right] = 
\frac{k_BT}{\chi_1} \left(\frac{\beta_p}{p^2}\right)
\label{eq:alphapss}
\eeq
where we have defined the matrix $\beta_p$ through the second equivalence (see 
eq.(\ref{eq:betap})).
Noting that
\beq
det(\alpha_p)=\left(\frac{k_BT}{\chi_1 p^2}\right)^3 
\left[1+\dot\gamma^2\left(\frac{\tau_1}{2p^2}\right)^2\right]
\eeq
the distribution function of the $p-th$ normal mode under steady shear flow 
results in eq.(\ref{eq:psipf}).
It is easy to check that eq.(\ref{eq:psipf}) reduces to the standard 
equilibrium distribution  
for $\dot\gamma=0$.

\section{Derivation of the sums}\label{app:sums}
In this appendix we outline the method to compute the sums in eqs. 
(\ref{eq:sumsq})
\beq
S_k(a,b)=\sum_{p=1}^\infty\frac{1}{p^{2+2k}}
\left[cos(p\pi a)-cos(p\pi b)\right]^2 \hskip 2cm k=0,1,2
\label{eq:b1}
\eeq
where $0\leq(a,b)\leq 1$. It is evident that $S_k(a,b)=S_k(b,a)\geq 0$ 
and the equality holds for $a=b$ only.
Developing the square inside the sum in eq.(\ref{eq:b1}) and using standard 
relations among trigonometric functions 
we obtain
\beq
S_k(a,b)=\sigma_k(0)+\frac{1}{2}
\left[\sigma_k(a)+\sigma_k(b)\right]-\sigma_k\left(\frac{a+b}{2}\right)
-\sigma_k\left(\frac{a-b}{2}\right)
\label{eq:b2}
\eeq
with
\beq
\sigma_k(x)=\sum_{p=1}^\infty\frac{cos(2\pi px)}{p^{2+2k}}
\label{eq:b3}
\eeq
so that the problem reduces to compute $\sigma_k(x)$ for $k=0,1,2$. 
Note that $\sigma_k(x)$ is an even and periodic function of $x$ with unitary 
period. 
Eq.(\ref{eq:b3}) expresses the Fourier series a function $f_k(x)$ with these 
properties, 
after removing the $p=0$ term.
Therefore we have to seek for a function $f_k(x)$ that satisfies the following 
conditions
\beqa
2p^{2+2k}\int_0^1~dx f_k(x)~cos(2\pi px)&=&1 \hskip 2cm p\geq1 \\ \nonumber
\int_0^1~dx f_k(x)~sin(2\pi px)&=&0
\label{eq:b4}
\eeqa
Applying repeatedly integration by parts we easily obtain the following 
identities
\beqa
\label{eq:b5}
\int_0^1~dx f_k(x)~cos(2\pi px)&=&
\frac{1}{(2\pi p)^2}\int_0^1~dx f'_k(x)~\partial_x cos(2\pi px) \\
&=&-\frac{1}{(2\pi p)^4}\int_0^1~dx f'''_k(x)~\partial_x cos(2\pi px) \\ \nonumber
&=&\frac{1}{(2\pi p)^6}\int_0^1~dx f^{\tt v}_k(x)~\partial_x cos(2\pi px)\nonumber
\eeqa
having imposed the boundary conditions ($f_k(0)=f_k(1)$) for the function 
and for its derivatives 
up to order five. 

Let us now consider the three $k$ values separately. For $k=0$ we have
\beqa
1&=&\frac{1}{2\pi^2}\int_0^1~dx f'_0(x)~\partial_x cos(2\pi px) \\ \nonumber
0&=&\frac{1}{(2\pi p)^2}
\int_0^1~dx f'_0(x)~\partial_x sin(2\pi px) \hskip 2cm p\geq 1  \\ \nonumber
f_0(0)&=&f_0(1)
\label{eq:b8}
\eeqa
where we have used the first of the equalities in eq.(\ref{eq:b5}) 
and a similar one for the sinus function.
These relations can be satisfied by choosing $f_0(x)=c_0+c_1~x+c_2~x^2$ 
with $c_0=0, c_1=-\pi^2$ and $c_2=\pi^2$
which provides $f_0(x)= \pi^2 (x^2-x)$. The function $\sigma_0(x)$ is then
\beq
\sigma_0(x)= f_0(x)-\int_0^1~dx~f_0(x)= \pi^2 (x^2-x-\frac{1}{6})
\label{eq:b9}
\eeq
and, after eq.(\ref{eq:b2})
\beq
S_0(a,b)=\frac{\pi^2}{2}|a-b|
\label{eq:b10}
\eeq
which is the wanted result for $k=0$.

For $k=1$ we must solve the problem
\beqa
1&=&-\frac{1}{8\pi^4}\int_0^1~dx f'''_1(x)~\partial_x cos(2\pi px) \\ \nonumber
0&=&-\frac{1}{(2\pi p)^4}
\int_0^1~dx f'''_1(x)~\partial_x sin(2\pi px) \hskip 2cm p\geq 1  \\ \nonumber
f_1(0)&=&f_1(1)
\label{eq:b11}
\eeqa
where the second equalities in eq.(\ref{eq:b5}) has been used. 
These relations can be satisfied choosing
$f'''_1(x)=c_3+c_4~x$ with $c_4=-8\pi^4$ and $c_3=4\pi^4$. 
Integrating three times over $x$ and imposing 
the boundary conditions on $f''_1, f'_1$ and $f_1$ we obtain 
$f_1(x)=\pi^4(2~x^3-x^2-x^4)/3$ which provides
\beq
\sigma_1(x)=f_1(x)-\int_0^1~dx~f_1(x)=\frac{\pi^4}{3}(\frac{1}{30}-x^2+2x^3-x^4)
\label{eq:b12}
\eeq
Using eq.(\ref{eq:b2}) we finally obtain
\beq
S_1(a,b)=\frac{\pi^4}{4}(a-b)^2
\left[(a+b)\left(1-\frac{a+b}{2}\right)-\frac{|a-b|}{3}\right]
\label{eq:b13}
\eeq

In the case $k=2$ the relation to be satisfied are 
\beqa
1&=&\frac{2}{(2\pi)^6}
\int_0^1~dx f^{\tt v}_1(x)~\partial_x cos(2\pi px) \\ \nonumber
0&=&\frac{1}{(2\pi p)^6}
\int_0^1~dx f^{\tt v}_1(x)~\partial_x sin(2\pi px) \hskip 2cm p\geq 1  \\ 
\nonumber
f_2(0)&=&f_2(1)
\label{eq:b14}
\eeqa
A procedure similar to the previous cases provides 
$f_2(x)=\pi^6 x^2(2x^4-6x^3+5x^2-1)/45$ and therefore
\beq
\sigma_2(x)=\frac{\pi^6}{45}\left[\frac{5}{7}+x^2(2x^4-6x^3+5x^2-1)\right]
\eeq
and
\beq
S_2(a,b)=\frac{\pi^6}{240}(a-b)^2
\left\{|a-b|^3+5(a+b)\left[(a+b)(2+a^2+b^2)-3a^2-2ab-3b^2\right]
\right\}
\eeq




\begin{table}
\caption{Equilibrium simulation details and results. $\tau_N$ is defined through 
the scaling
laws given in the text. $N_T$ is the total length of the simulation run in time 
steps.
$T/\tau_N$ is the total simulation time over the longest 
chain relaxation 
time.}
\label{table1}
\begin{tabular}{ccccccc}
 $N$ &$N_T$x$10^{-6}$& $R_g$   & $R$    & $\tau_N$ & $T/\tau_N$ \\
\tableline
  9  &   2      & 1.413(5)& 3.44(1)& 3.04& 165 \\
 20  &  10      & 2.340(2)& 5.83(1)& 19.1& 131 \\ 
 30  &  50      & 3.07(2) & 7.77(5)& 47.6& 263 \\ 
 50  &  50      & 4.16(1) & 10.4(1)& 149 &  84 \\ 
100  & 300      & 6.410(1)& 16.27(2)&677 & 111 \\
\end{tabular}
\end{table}
\newpage

\begin{table}
\caption{Flow simulations details and results for the stiff spring chain model
without EV interactions.
$N=Rouse$ means data for the continuous model of section \ref{sec:rouse}. 
$T/\tau_N$
is the total time of the simulation over the equilibrium longest relaxation 
time of the chain}
\label{table2}
\begin{tabular}{ccccccccc}
$N$ &$\beta$&$T/\tau_N$& $m_O$ & $m_G$ & $\delta G_I$&$\delta G_{II}$&$\delta 
G_{III}$\\
\tableline
100 &  3.2  &  200     & 2.8(3)&1.7(1)  & 2.6(3)     &  -0.33(1)     & -
0.09(1)\\
300 &  3.2  &  120     & 2.7(3)&1.78(4) & 5.4(1.2)   &  -0.32(2)     & -
0.02(1)\\
Rouse& 3.2  &  -       & 2.5   &1.75    & 5.00828    &  -0.32714     &  0      
\\
\tableline
100  & 10.0 &  200     & 3.7(1)&2.25(3) & 21(2)      &  -0.44(1)     & -0.18(1) 
\\
300  & 10.0 &  60      & 2.8(2)&1.97(6) & 35(3)      &  -0.41(2)     & -0.03(2) 
\\
Rouse& 10.0 &  -       & 2.5   &1.75    & 46.06164   &  -0.34736     & 0 \\
\end{tabular}
\end{table}

\begin{table}
\caption{Flow simulations details and results for the stiff spring chain model 
with EV interactions.
Symbols are like in table II. Data from ref.[38] are also reported for
comparison.}
\label{table3}
\begin{tabular}{cccccccccc}
$N$ &$\beta$&$T/\tau_N$& $m_O$ & $m_G$ & $\delta G_I$&$\delta G_{II}$&$\delta 
G_{III}$\\
\tableline
\tableline
20 & 3.2  &  131& 4.2(1)&2.60(6) & 1.41(3)    &  -0.430(6)    & -0.0061(2)\\
30 & 3.2  &  263& 4.0(1)&2.5(2)  & 1.5(1)     &  -0.42(1)     & -0.16(2)  \\
50 & 3.2  &  168& 3.7(4)&2.5(2)  & 1.8(2)     &  -0.40(3)     & -0.17(1)  \\
100& 3.2  &  111& 4.1(6)&2.3(3)  & 1.4(1)     &  -0.40(1)     & -0.21(2)  \\
300& 3.2  &   98& 3.3(3)&2.1(3)  & 2.5(3)     &  -0.47(2)     & -0.14(2)  \\
\tableline
50\tablenotemark[1]& 
     3.2  &  73       & 4.6(7)&2.1(9)  & 1.8(2)     &  -0.47(1)     & -0.12(1)  
\\
\tableline
\tableline
100& 10.0 &   75      & 6.2(3)&3.9(1)  & 5.2(2)     &  -0.60(4)     & -0.34(5)  
\\
300& 10.0 &   59      & 5.2(2)&3.4(1)  & 8.1(8)     &  -0.57(1)     & -0.24(2) 
\\
\tableline
 30\tablenotemark[1]&
     10.0 &  167      & 6.2(4)&3.4(3)  & 5.3(3)     &  -0.62(3)     & -0.40(4) 
\\
\end{tabular}
\tablenotetext[1]{data from ref.\cite{PR95}}
\end{table}

\vfill
\eject

\vfill
\eject

\vfill
\eject

\vfill
\eject

\vfill
\eject

\vfill
\eject

\vfill
\eject

\vfill
\eject

\vfill
\eject

\vfill
\eject

\vfill
\eject

\newpage
\noindent Corresponding author: \\
Carlo Pierleoni \\
Physics Department, University of L'Aquila \\
Via Vetoio, Coppito 67010-L'Aquila (Italy) \\
phone: +39+0862+433056 \\
fax:~~~+39+0862+433033 \\
email: carlo.pierleoni@aquila.infn.it

\hfill
\eject


\begin{thebibliography}{10}

\bibitem{Y71}
H. Yamakawa {\em Modern Theory of Polymer Solutions}, Harper \& Row (New York, 
1971).

\bibitem{Ge79}
P.-G. de~Gennes, {\em {Scaling Concepts in Polymer Physics}} (Cornell
  University Press, Ithaca,N.Y., 1979).

\bibitem{DE86}
M. Doi and S. F. Edwards, {\em The Theory of Polymer Dynamics}, Clarendon Press,
Oxford (1986).

\bibitem{dCJ90}
J. des Cloizeaux and G.Jannink {\em Polymers in Solution: Their Modeling and 
Structure},
Clarendon Press (Oxford 1990).

\bibitem{O96}
H. C. \"Ottinger {\em Stochastic Processes in Polymeric Fluids}, Springer 
(Berlin 1996).

\bibitem{PSO86}
S. Puri, B. Schaub and Y. Oono, Phys. Rev. {\bf A34}, 3362 (1986).

\bibitem{W89}
S.Q. Wang, Phys. Rev. {\bf A40}, 2137 (1989).

\bibitem{O90}
H.C. \"Ottinger, Phys. Rev. {\bf A41}, 4413 (1990).

\bibitem{W90}
S.Q. Wang, J. Chem. Phys. {\bf 92}, 7618 (1990).

\bibitem{ZO91}
W. Zylka and H.C. \"Ottinger, Macromolecules {\bf 24}, 484 (1991).

\bibitem{BH90}
P.R. Baldwin and E. Helfand, Phys. Rev. {\bf A41}, 6772 (1990).

\bibitem{RK89}
Y. Rabin and K. Kawasaki, Phys. Rev. Letts.{\bf 62}, 2281 (1989).

\bibitem{Carl95-96}
W. Carl, Macromol. Theory Simul. {\bf 4}, 77 (1995), {\em ibid.} {\bf 5}, 1 
(1996).

\bibitem{CMS69}
F.R. Cottrell, E.W.Merrill and K.A.Smith, J. Polym. Sci. PartA-2 {\bf 7}, 1415 
(1969).

\bibitem{LO88}
P. Lindner and R.C. Oberth\"ur, Colloid Polym. Sci. {\bf 266}, 886 (1988), 
Physica 
{\bf B 156\&157}, 410 (1989)

\bibitem{Li91}
P. Lindner,  in {\em {Neutron, X-ray and Light Scattering}}, edited by P.
  Lindner and T. Zemb (Elsevier science Publishers, 1991).

\bibitem{LS93}
A. Link and J. Springer, Macromolecules {\bf 26}, 464 (1993).

\bibitem{ZS94-95}
M. Zizenis and J. Springer, Polymer {\bf 35}, 3156 (1994), {\bf 36}, 3459 
(1995).

\bibitem{LSM97}
E.C.Lee, M.J.Solomon and S.J.Muller, Macromolecules {\bf 30}, 7313 (1997).

\bibitem{LM99}
E.C.Lee and S.J.Muller, Polymer {\bf 40}, 2501 (1999).

\bibitem{BSF92}
S. Smith, L. Finzi, and C. Bustamante, Science {\bf 258},  1122  (1992).

\bibitem{BMSS94}
C. Bustamante, J.F. Marko, E.D. Siggia and S. Smith, Science {\bf 265}, 1599 
(1994).

\bibitem{CLHLVCC96}
P Cluzel, A. Lebrun, C. Heller, R. Lavery, J.-L. Viovy, D. Chatenay and F. 
Caron,
Science {\bf 271}, 729 (1996).

\bibitem{ROH97}
M.~Rief, F.~Oesterhelt, B.~Heymann, and H.~E.~Gaub,
\newblock Science {\bf 275}, 1295 (1997).

\bibitem{PSC94}
T.T. Perkins, D.E. Smith and S. Chu, Science {\bf 264}, 819 (1994).

\bibitem{PSLC95}
T.T. Perkins, D.E. Smith, R.G. Larson and S.Chu, Science {\bf 268}, 83 (1995).

\bibitem{PQDC94}
T.T. Perkins, S.R. Quake, D.E. Douglas and S. Chu, Science {\bf 264}, 822 
(1994).

\bibitem{PSC97}
T.T. Perkins, D.E. Smith and S. Chu, Science {\bf 276}, 2016 (1997)

\bibitem{SBC99}
 D.E. Smith, H.P. Babcock,S. Chu, Science, {\bf 283},1724,(1999)

\bibitem{LHBW99}
P.LeDuc, C.Haber, G. Bao and D. Wirtz, Nature {\bf 399}, 564 (1999).

\bibitem{Liu89}
T.W. Liu, J. Chem Phys. {\bf 90}, 5826 (1989).

\bibitem{Carl}
W. Carl and W. Bruns, Macromol. Theory Simul. {\bf 3}, 295 (1994).

\bibitem{Liulyn}
A.V. Lyulin, D. Adolf and G.R. Davies, J. Chem. Phys. {\bf 111}, 758 (1999).

\bibitem{ADMH98}
N.C. Andrews, A.K. Doufas and A.J. McHugh, Macromolecules {\bf 31}, 3104 (1998).

\bibitem{RKKZ99}
R.Rzehak, D. Kienle, T. Kawakatsu and W. Zimmermann, Europhys. Letts. {\bf 46}, 
821 (1999).

\bibitem{LL00}
L. Li and R.G. Larson, Macromolecules{\bf 33},1411(2000) 

\bibitem{PR93}
C. Pierleoni and J.-P. Ryckaert, Phys. Rev. Letts. {\bf 71}, 1724 (1993).

\bibitem{PR95}
C. Pierleoni and J.-P. Ryckaert, Macromolecules {\bf 28}, 5097 (1995).

\bibitem{RP99}
J.-P. Ryckaert and C. Pierleoni, in 
{\em Flexible Polymer Chain Dynamics in Elongational Flow: Theory and 
Experiment}, 
T.Q. Nguyen and H.H. Kausch eds., Springer-Verlag (Berlin, 1999).

\bibitem{AKH99}
C. Aust, M. Kr\"oger and S. Hess, Macromolecules {\bf 32}, 5660 (1999).

\bibitem{DR}
E. Duering and Y. Rabin, Macromolecules {\bf 23}, 2232 (1990), J. Rheology {\bf 
35}, 213 (1991).

\bibitem{Lai}
P.Y Lai, in {\em Flow-Induced Structure in Polymers}, A.I. Nakatami and M.D. 
Dadmun eds., ACS Symposium 
{\bf 597}, (1995).

\bibitem{P76}
P. Pincus, Macromolecules {\bf 9},  386  (1976).

\bibitem{FBC78}
B. Farnoux {\it et~al.}, J. Physique {\bf 39},  77  (1978).

\bibitem{PAR97}
C. Pierleoni, G. Arialdi and J.-P. Ryckaert, Phys. Rev. Letts. {\bf 79}, 2990 
(1997).

\bibitem{On85}
A. Onuki, J. of Phys. Soc. Japan {\bf 54},  3656  (1985).

\bibitem{BDODCFJ75}
H. Benoit {\it et~al.}, Macromolecules {\bf 8},  451  (1975).

\bibitem{PB88}
N. Pistoor and K. Binder, Colloid \& Polymer Sci. {\bf 266}, 132 (1988).

\bibitem{BCAH87}
R. B. Bird, C. F. Curtiss, R. C. Armstrong and O. Hassager, {\em Dynamics of 
Polymeric Liquids}
2nd ed., Vol. 2: Kinetic Theory, John Wiley \& Sons (New York 1987).

\bibitem{NumRecF}
W.H. Press, S.A. Teukolsky, W.T. Vetterling and B.P. Flannery, {\em Numerical 
Recipes}, 
Cambridge University Press (New York, 1992).

\bibitem{On97}
A. Onuki, J. Phys.: Cond.Mat. {\bf 9}, 6119 (1997).

\bibitem{CR86}
G. Ciccotti and J.-P. Ryckaert, Comp. Phys. Rep. {\bf 4}, 345 (1986).

\bibitem{AT86}
M.P. Allen and D.J. Tildesley, {\em The Computer Simulation of Liquids}, 
Clarendon (1987).

\bibitem{PR91-92}
C. Pierleoni and J.-P. Ryckaert, Phys. Rev. Letts. {\bf 66}, 2992 (1991), 
J. Chem. Phys. {\bf 96}, 3574 (1992).

\bibitem{DK91-93}
B. D\"unweg and K. Kremer, Phys. Rev. Letts. {\bf 66},2996 (1991), 
J. Chem. Phys. {\bf 99}, 6983 (1993).

\bibitem{Z69}
R. Zwanzig, Adv. Chem. Phys. {\bf 15}, 325 (1969).



\end{thebibliography}
\end{document}